\documentclass[preprint,12pt,authoryear]{elsarticle}

\usepackage{lineno}
\usepackage{amssymb}
\usepackage{amsmath}
\usepackage[ruled,vlined,linesnumbered]{algorithm2e}
\usepackage{subfigure}
\usepackage{threeparttable}
\usepackage{tabularx}
\usepackage{soul}
\usepackage{lineno}
\usepackage{hyperref}
\usepackage[nameinlink]{cleveref}
\Crefname{figure}{{Fig.}}{{Figs.}}
\Crefname{equation}{{}}{{}}
\Crefname{section}{Section}{Sections}
\usepackage{enumitem}
\usepackage{fancyhdr}
\usepackage{booktabs}
\hypersetup{
  colorlinks=true,
  citecolor=blue,
  linkcolor=blue,
  urlcolor=blue}
\usepackage{caption}

\journal{Expert Systems With Applications }

\begin{document}

\begin{frontmatter}



\title{Geomagnetic and Inertial Combined Navigation Approach Based on Flexible Correction-Model Predictive Control Algorithm} %% Article title


\author[1]{Xiaohui Zhang \corref{cor1}} %% Author name
\author[1]{Xingming Li} 
\author[1]{Songnan Yang \corref{cor2}} 
\author[1]{Wenqi Bai} 
\author[1]{Yirong Lan} 

\affiliation[1]{organization={School of Automation and Information Engineering},%Department and Organization
            addressline={Xi’an University of Technology}, 
            city={Xi'an},
            postcode={710048}, 
            state={Shannxi},
            country={China}}
\cortext[cor1]{This work was supported in part by the National Major Scientific Instrument Development Project of China under Grant 62127809; in part by the National Natural Science Foundation of China under Grant 62073258; and in part by the Basic Research in Natural Science and Enterprise Joint Foundation of Shaanxi Province under Grant 2021JLM-58.}
\cortext[cor2]{Corresponding author}

%% Abstract
\begin{abstract}
This paper proposes a geomagnetic and inertial combined navigation approach based on the flexible correction-model predictive control algorithm (Fc-MPC). This approach aims to overcome the limitations of existing combined navigation methods that require prior geomagnetic maps and the inertial navigation drift of long-range missions. The proposed method uses geomagnetic gradient information and the model predictive control (MPC) algorithm with heading control and state constraints, eliminating the dependence on prior geomagnetic maps. Instead, the proposed method achieves real-time measurements of the geomagnetic declination, geomagnetic inclination, and inertial navigation data and introduces uniform compensation conditions to adjust and correct the predictive results in real-time. Simulation and real experiment results demonstrate that the proposed Fc-MPC algorithm significantly improves the precision, efficiency, and stability of the geomagnetic and inertial combined navigation system.
\end{abstract}

\begin{graphicalabstract}
\includegraphics[width=0.7\linewidth]{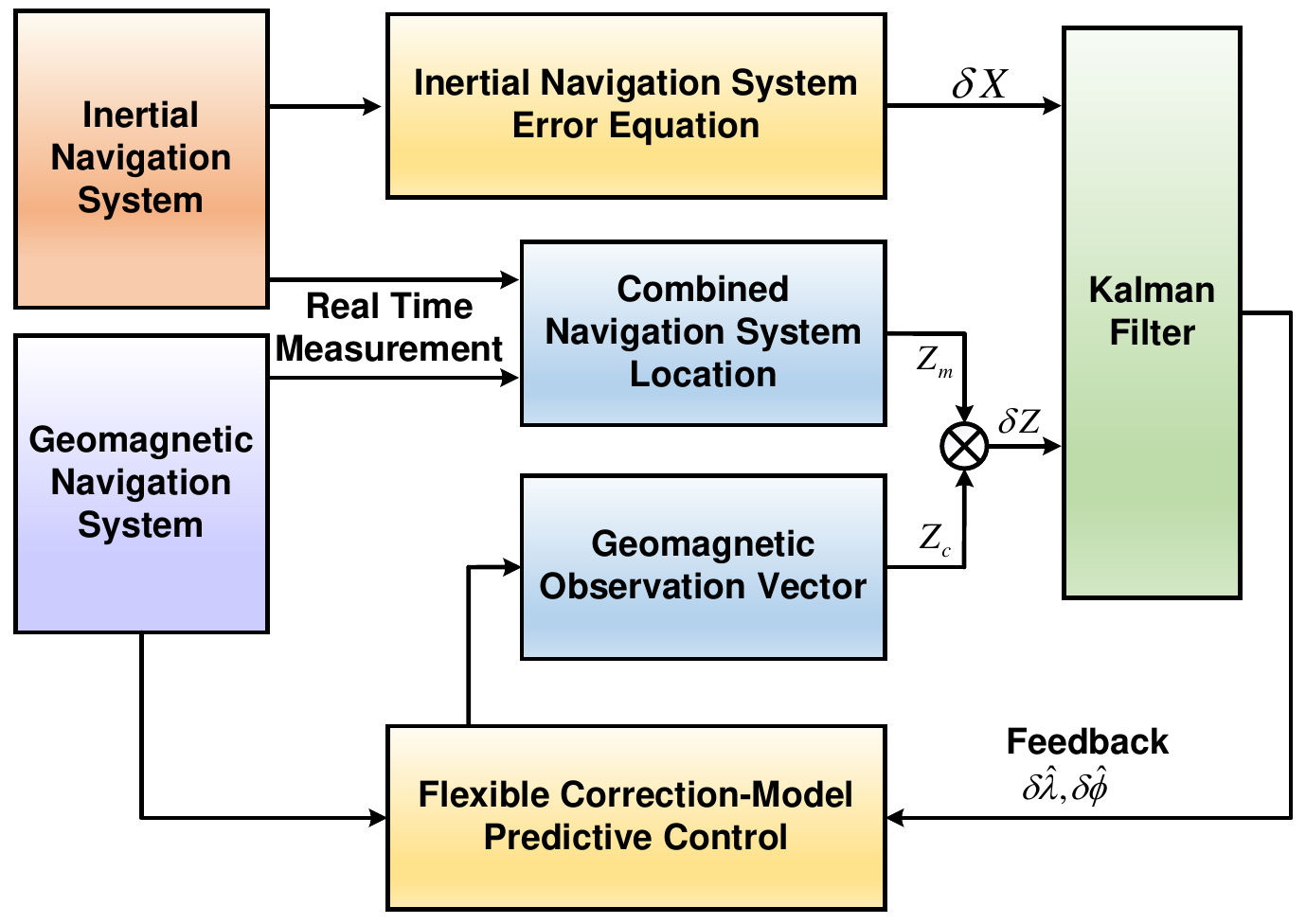}
\end{graphicalabstract}

\begin{highlights}
\item Hybrid navigation method combining geomagnetic and inertial systems for long-range navigation in GPS system-denied areas.
\item Flexible heading correction method addressing sensitivity to real-time information errors of model predictive control in navigation systems.
\item Achieves high navigation accuracy and effectively resists unknown magnetic storms while correcting accumulated drift errors.
\item The method demonstrates strong performance by utilizing the World Magnetic Model, real magnetic storm data, and actual measurement.
\end{highlights}

%% Keywords

\end{frontmatter}

 %\linenumbers
\section{Introduction}
\label{Section:1}
In the current context of increasing international tensions, passive navigation methods—such as geomagnetic, inertial, and visual systems are emerging as reliable alternatives for navigation in GPS-denied environments. This shift is gradually gaining attention from researchers \citep{GarciaPulido2017,Chang2023, Wang2023}. The geomagnetic navigation system is capable of long-distance navigation while excelling in security and stealth \citep{Goldenberg2006}. However, it encounters challenges in regions with low geomagnetic gradients or significant geomagnetic interference, which can affect precision and stability. In contrast, the inertial navigation system (INS) provides high accuracy over short durations and is less vulnerable to external variations \citep{Engelsman2023}. Nonetheless, as time progresses, cumulative errors in the inertial system tend to increase. The visual navigation system can achieve exceptionally high accuracy once the destination is locked and can adjust the course if the destination moves \citep{Zhao2023}. However, for long-distance missions, visual navigation requires support from other systems before identifying the destination due to constraints imposed by target characteristics and environmental changes. Research on navigation systems utilizing these methods is limited to single navigation data, which hinders efforts to further improve the accuracy and reliability of the navigation system.

These limitations can be addressed by combining the geomagnetic navigation system with an inertial navigation system \citep{Pang2023}. Inertial navigation benefits from its independence from external factors like weather and terrain. Although the inertial navigation system has the error accumulation problem, geomagnetic navigation systems can effectively mitigate these shortcomings \citep{Qi2023}. The integration of data from geomagnetic and inertial navigation can enhance the overall precision of the navigation system. The geomagnetic and inertial combined navigation system has benefits such as solid concealment, robust resistance to interference, global applicability, and the elimination of cumulative errors \citep{Zhao2024}. Researchers have conducted extensive studies on the inertial and geomagnetic combined navigation method.
\citet{Liu2007} proposed an improved correlation matching algorithm based on contour constraints to improve the accuracy of geomagnetic-aided combined navigation.
\citet{Duan2019} proposed a combined navigation method built upon matching strategy and hierarchical filtering to improve the positioning precision of the inertial and geomagnetic combined navigation system algorithm and improved the probability data association-iterated closest contour point (PDA-ICCP) geomagnetic matching algorithm. \citet{Ding2022} proposed a two-stage combined matching algorithm, including coarse matching and contour-based fine matching, to improve the precision and robustness of underwater terrain-aided navigation system.
 \citet{Xu2020} proposed an improved iterated closest contour point (ICCP) algorithm to help eliminate errors from the INS. \citet{Wang2022} proposed a inertial and geomagnetic combined matching navigation algorithm for long-range navigation. This method integrates the map-based geomagnetic matching navigation method with the geomagnetic bionic navigation method.
\citet{Pang2023} proposed an adaptive geomagnetic navigation method. It roughly estimates navigation errors based on the error characteristics of the inertial navigation system, then performs geomagnetic matching for positioning to improve matching accuracy and timeliness. However, these methods rely on accurate a priori geomagnetic maps, which are often difficult to obtain in practical navigation tasks. This limitation restricts the utilization of the combined navigation system \citep{Wang2020}. \citet{ZhangYi2021} introduced the model predictive control (MPC) algorithm into long-distance geomagnetic navigation for the first time to solve the problem of no prior geomagnetic map. Since this method relies solely on geomagnetic information, the accuracy of the Earth's magnetic field and heading calculation model will directly influence the method's overall precision. Additionally, this approach does not incorporate a course correction mechanism, and the affects of magnetic storms on its performance remain uncertain. In the face of complex and unknown GPS-denied environments, the existing geomagnetic and inertial combined navigation methods \citep{Scott2011,Rong2020} exhibit a limitation in the dynamic model predictive updates necessary for effectively addressing geomagnetic interference. These limited the use of the combined navigation system \citep{Teng2018,Wu2024}. Summarizing the current research, combined navigation methods continue to encounter several pressing challenges:
\begin{enumerate}
    \item Geomagnetic navigation methods depend on precise a priori geomagnetic maps, which are often challenging to acquire for practical navigation tasks. This limitation renders them ineffective in environments lacking prior geomagnetic maps, limiting the applicability of combined navigation systems.
    \item MPC algorithms in current research depend on precise mathematical models to predict heading angles. This reliance renders existing methods highly sensitive to external disturbances, resulting in diminished accuracy and reliability when real-world conditions deviate from the model. Consequently, the practical application of integrated navigation systems is constrained.
    \item Existing geomagnetic and inertial combined navigation methods lack a dynamic prediction update mechanism for multi-parameter models when confronted with complex, unknown time varying interference. This deficiency leads to inadequate anti-interference performance. These issues need to be addressed urgently to improve the reliability and practicality of combined navigation systems in complex environments.
\end{enumerate}

\begin{figure}[htb]
    \centering
    \includegraphics[width=0.55\linewidth]{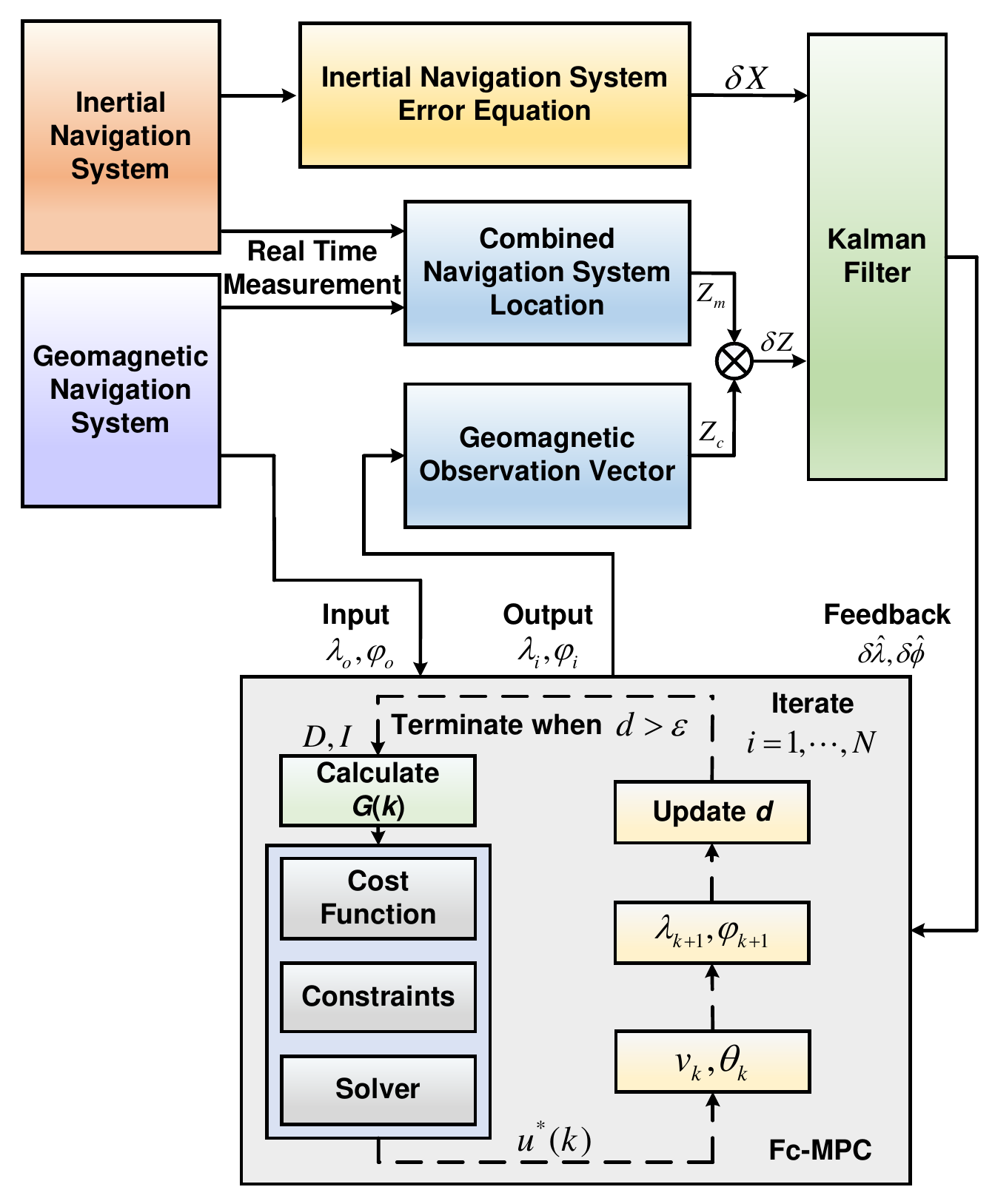}
    \caption{The flowchart of flexible correction-model predictive control algorithm for combined navigation systems.}
    \label{fig:1}
\end{figure}

To address these challenges, this paper introduces the  Fc-MPC algorithm for the geomagnetic and inertial combined navigation system. This approach leverages geomagnetic gradient information and the MPC algorithm with heading control and state constraints, without relying on geomagnetic maps or prior geographic data. Navigation is achieved solely through measurements of magnetic declination, magnetic inclination, and inertial navigation data. A uniform compensation condition is introduced for real-time predictive correction. The Kalman filter integrates inertial navigation errors with position errors from geomagnetic navigation over short time intervals, enhancing the accuracy of the combined navigation approach. To prevent the accumulation of long-term inertial navigation errors, the algorithm considers only the state variables at the next sampling moment when the geomagnetic gradient matrix remains unchanged. The proposed Fc-MPC algorithm flowchart is shown in \Cref{fig:1}. The contributions of this paper can be summarized as follows:

\begin{enumerate}
    \item To address the challenges of geomagnetic and inertial navigation in GPS-denied environments without prior geomagnetic maps, this paper proposes the Fc-MPC method, which utilizes combined real-time geomagnetic and inertial data for heading correction.
    
    \item This paper presents a flexible heading correction method designed to address the vulnerabilities of traditional data fusion algorithms to real-time information errors caused by magnetic storms in combined navigation systems.
    
    \item This paper overlays a geomagnetic anomaly onto the World Magnetic Model model to assess and verify the stability of the combined navigation approach based on the algorithm, specifically the algorithm addressing geomagnetic susceptibility under time-varying interference. Additionally, this algorithm is applied to real navigation systems.
\end{enumerate}

The remainder of this article is structured as follows: \Cref{Section:2} introduces the basic theory of geomagnetic navigation and the MPC algorithm for navigation systems. \Cref{Section:3} proposes a geomagnetic and inertial combined navigation approach based on the Fc-MPC algorithm. \Cref{Section:4} analyzes and evaluates the performance of the method through various interference simulations and experiments with real data. Finally, \Cref{Section:5} concludes this work.

\section{Basic Theory}
\label{Section:2}
\subsection{Geomagnetic Field Model}
\label{Section:2.1}
\begin{figure}[h]
    \centering
    \includegraphics[width=0.45\linewidth]{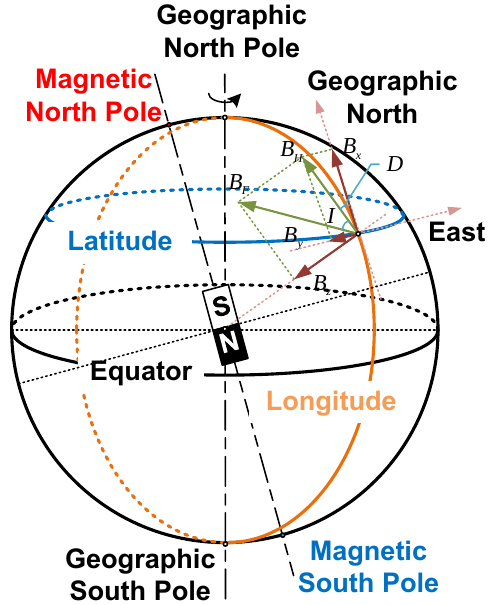}
    \caption{Global Geomagnetic Field Model.}
    \label{fig:2}
\end{figure}

Constructing an accurate geomagnetic model becomes the basis for achieving geomagnetic navigation. The spherical harmonic analysis method is currently the most commonly used method for constructing global geomagnetic field models \citep{Kaji2019}. The international geomagnetic reference field (IGRF) model and the WMM model \citep{Guo2013,Zhangjia2021} commonly used by researchers are both based on the spherical harmonic analysis method. The geomagnetic field is characterized by an axial dipole field aligned with the geocentric axis, as shown in \Cref{fig:2}.

The geomagnetic navigation system can determine the location through the uniqueness of geomagnetic information. During navigation, the location of the observation point can be determined by the description of geomagnetic information as shown in Eq.\Cref{eq:1}:
\begin{align}
\textit{\textbf{B}} = \{ {B_x},{B_y},{B_z},{B_H},{B_F},D,I\} .
\label{eq:1}   
\end{align}

The geomagnetic information vector \textbf{\textit{B}} is composed of seven elements, specifically: \(B_x\),\(B_y\),\(B_z\) represents the geomagnetic field components toward the north, east, and radially inward, respectively; \(B_H\) represents the horizontal intensity; \(B_F\) represents the total geomagnetic field intensity; \textit{D} represents the magnetic declination; \textit{I} represents the magnetic inclination. The magnetic declination \textit{D} represents the angle formed between the horizontal intensity \(B_H\) and the geographic north direction \textit{x}. The magnetic inclination \textit{I} represents the angle formed between the total intensity \(B_F\) and the horizontal plane \textit{xoy}. Based on the definitions of magnetic declination and magnetic inclination, the magnetic declination \textit{D} and magnetic inclination \textit{I} can determine a unique location. The specific calculations are as follows:
\begin{align}
\left\{ \begin{array}{l}
D = \arctan ({B_y}/{B_x})\\
I = \arctan ({B_z}/\sqrt {B_x^2 + B_y^2}).
\end{array} \right.
\label{eq:2}
\end{align}

\subsection{Mapless Geomagnetic Navigation Method}
\label{Section:2.2}
\begin{figure}[htp]
    \centering
    \includegraphics[width=0.75\linewidth]{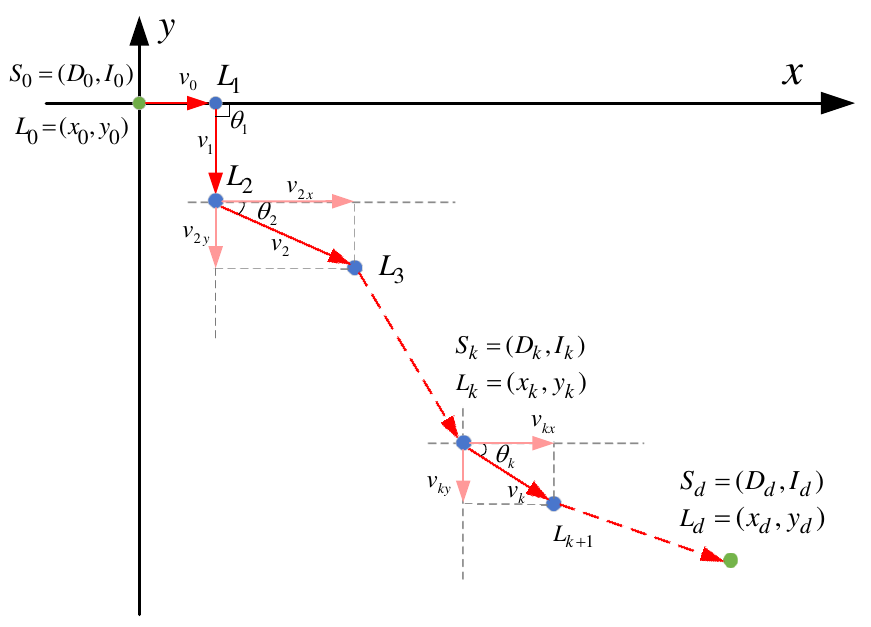}
    \caption{Schematic diagram of the mapless geomagnetic navigation method by using geomagnetic declination and inclination.}
    \label{fig:3}
\end{figure}

This section introduces how to realize mapless geomagnetic navigation by utilizing two geomagnetic elements from the geomagnetic field model, as shown in \Cref{fig:3}. 

By analyzing the properties of the geomagnetic field, a specific location can be identified using geomagnetic declination \textit{D} and inclination \textit{I} \citep{Zhangjia2021}. Assuming \({L(k)}={({x_k},{y_k})^T}\)represents the sampling location at time \textit{k}, where \(x_k\) and \(y_k\) denote the coordinates in the \textit{xoy} coordinate system. At the same time, the geomagnetic navigation calculates longitude \(\lambda _c\) and latitude \(\varphi _c\) can be obtained based on existing methods \citep{Ziggah2016,Civicioglu2012}. The state variables at time \textit{k} are \textit{D} and \textit{I}, denoted as \({S_k}={({D_k},{I_k})^T}\). \(L_0\) and \(L_d\) denote the positions of the starting point and the target point, respectively. \(S_0\) and \(S_d\) denote the states of the starting point and target point, respectively. The \textit{k}-th time heading can be calculated using Eq.\Cref{eq:3} and Eq.\Cref{eq:4}:
 \begin{align}
{v_k}=\sqrt {v_{xk}^2 + v_{yk}^2},
\label{eq:3}
\end{align} 
 \begin{align}
\theta_k = \pm  \arctan({v_{yk}}/{v_{xk}}),
\label{eq:4}
\end{align}  
where \(v_k\) represent the velocity at time \textit{k}. \(\theta_k\) represent the angle formed between the velocity \(v_k\) and the positive \textit{x} direction. 
\(v_{xk}\) and \(v_{yk}\) represent the components of the vector \(v_k\) in the positive \textit{x} direction the positive \textit{y} direction respectively \citep{Chen2020,Chen2022}.

The gradients of \textit{D} and \textit{I} at time \textit{k} are denoted by \(G(k)\), which can be calculated as Eq.\Cref{eq:5}:
 \begin{align}
 G(k)=
\begin{pmatrix}
    g_{Dkx} & g_{Dky} \\
    g_{Ikx} & g_{Iky}
\end{pmatrix},
\label{eq:5}
\end{align}
where \(g_{Dkx}\), \(g_{Dky}\) , \(g_{Ikx}\) and \(g_{Iky}\) represent the gradients of \textit{D} in the \textit{x} and \textit{y} directions, and the gradients of \textit{I} in the \textit{x} and \textit{y} directions at time \textit{k} respectively. The gradients of  \textit{D} and \textit{I} at \(L_2\) are shown in Eq.\Cref{eq:6} to Eq.\Cref{eq:9}:
 \begin{align}
 g_{D2x}=(D_1-D_0)/(x_1-x_0),
\label{eq:6}
\end{align} 
 \begin{align}
 g_{D2y}=(D_2-D_1)/(y_2-y_1),
\label{eq:7}
\end{align} 
 \begin{align}
 g_{I2x}=(I_1-I_0)/(x_1-x_0),
\label{eq:8}
\end{align} 
 \begin{align}
 g_{I2y}=(I_2-I_1)/(y_2-y_1).
\label{eq:9}
\end{align} 

Obtaining the gradient at each sampling location, along with the additional distance from each step to the previous one, will reduce the navigation real-time calculation performance. Therefore, this paper only calculates the gradient at the starting point. At each sampling position, the gradient is updated using \(\Delta D_{(k+1)}\) and \(\Delta I_{(k+1)}\) (the changes in \textit{D} and \textit{I}), the heading angle \(\theta_k\), the velocities \(v_{xk}\) and \(v_{yk}\) \citep{Wang2018a,Zhou2022}. The gradient at the sampling position at time (\textit{k}+1) can be approximately updated as shown in Eq.\Cref{eq:10} to Eq.\Cref{eq:12} \citep{Guo2019,ZhangYi2021}:
 \begin{align}
 G(k+1)=G(k)+
\begin{pmatrix}
    \Delta D_{k+1} \\
    \Delta I_{k+1}
\end{pmatrix} 
\begin{pmatrix}
    \frac {\cos\theta_k} {v_{xk}T}  &
    \frac {\sin\theta_k} {v_{yk}T}
\end{pmatrix}, 
\label{eq:10}
\end{align}
 \begin{align}
\Delta D_{k+1}=D_{k+1}-D_k,
\label{eq:11}
\end{align} 
 \begin{align}
\Delta I_{k+1}=I_{k+1}-I_k.
\label{eq:12}
\end{align} 

%%%%%%%%%%%%%%%%%%%%%%%%%%%%%%%%%%%%%%%%%%%%%%%%%%%%%%%%%%%
\subsection{Model Predictive Control for Navigation}
\label{Section:2.3}
This section introduces the MPC algorithm for geomagnetic navigation by utilizing the mapless geomagnetic navigation method theory. 

The model input is denoted as $U(k) = {({v_{xk}},{v_{yk}})^T}$, where \(v_{xk}\) and \( v_{yk} \) represent the velocity components at time \( k \) along the \textit{x} and \textit{y} directions, respectively. Wherein the magnetic declination \textit{D} and the magnetic inclination \textit{I} are designated as state variables $S(k) = {({D_k},{I_k})^T}$. Specifically, \(D_k\) denotes the magnetic declination at time \textit{k}, while \(I_k\) represents the magnetic inclination at time \textit{k}. The uncertainty model is established on the zero-order holder with a specified sampling period of \textit{T} and transformed into the following discrete form as follows \citep{Choi2022}:

\begin{align}\begin{aligned}
    S(k + 1) = AS(k) + B(k)U(k),
\end{aligned}
\label{eq:13}
\end{align} 
\begin{align}
A=I_2,
\label{eq:14}
\end{align} 
\begin{align}
B(k)=G(k)T,
\label{eq:15}
\end{align} 

Prediction is performed over a prediction interval of a given length of \textit{N}. For the convenience of subsequent calculations, the predicted vectors at each time step are calculated as follows: 
\begin{align}
\bar{S}(k+1)=
\begin{pmatrix}
    S(k+1|k) \\
    S(k+2|k) \\
    \vdots \\
    S(k+N|k) 
\end{pmatrix}, 
\label{eq:16}
\end{align} 
\begin{align}
\bar{U}(k+1)=
\begin{pmatrix}
    U(k|k) \\
    U(k+1|k) \\
    \vdots \\
    U(k+N-1|k) 
\end{pmatrix}, 
\label{eq:17}
\end{align} 
where $U(k + i|k)$ and $S(k + i|k)$ represent the system input and state vectors predicted at time \textit{k}, where $i = 1,2, \cdots,N$.

The geomagnetic values of magnetic declination and magnetic inclination do not change significantly within \(1^\circ\)  (approximately 111.32 km) \citep{Gavoret1986}. Therefore, $G(k)$ and $B(k)$ are considered constant within the prediction interval. The initial conditions at time \textit{k} is as follows: 
\begin{align}
\left\{ \begin{array}{l}
S(k|k)=S(k)\\
U(k|k)=U(k).
\end{array} \right.
\label{eq:18}
\end{align}

The state $S(k + i|k)$ predicted at time \textit{k} is as follows:
\begin{align}
\left\{\!\!\!\begin{array}{l}
S\!(k+1|k)=AS\!(k|k)+BU\!(k|k)\\
S\!(k+2|k)=AS\!(k+1|k)+BU\!(k+1|k)\\
\vdots \\
S\!(k{+}N|k){=}AS\!(k{+}N{-}1|k){+}BU\!(k{+}N{-}1|k),
\end{array} \!\!\!\right.
\label{eq:19}
\end{align}
where, the expression for $\bar S(k)$ and $\bar U(k)$ are as follows:
\begin{align}
\bar{S}(k+1)=MS(k)+C(k)\bar{U}(k),
\label{eq:20}
\end{align} 
\begin{align}
M=
\begin{pmatrix}
    A & A^2 & \cdots & A^N
\end{pmatrix} ^T\!\!,
\label{eq:21}
\end{align} 
\begin{align}\begin{aligned}
C(k)=
\begin{pmatrix}
    B(k) & 0 & \cdots & 0 \\
    AB(k) & B(k) & \cdots & 0 \\
    \vdots & \vdots & \quad &\vdots \\
    A^{N-1}B(k) &  A^{N-2}B(k) & \cdots &  B(k) 
\end{pmatrix}
\\= 
\begin{pmatrix}
    G(k)T & 0 & \cdots & 0 \\
    AG(k)T & G(k)T & \cdots & 0 \\
    \vdots & \vdots & \quad &\vdots \\
    A^{N-1}G(k)T &  A^{N-2}G(k)T & \cdots &  G(k)T
\end{pmatrix}.
\label{eq:22}
\end{aligned}
\end{align} 

The loss function is defined as follows:
\begin{align}\begin{aligned}
\begin{array}{l}
J(k) = \sum\limits_{j = 0}^{N - 1} {U{{(k + j|k)}^T}RU(k + j|k)} \\
 {+} \sum\limits_{i {=} 1}^{N {-} 1} {(S(k {+} i|k)} {-} {S_d}{)^T}Q(S(k {+} i|k) {-} {S_d})\\
 {+} {(S(k {+} N|k) {-} {S_d})^T}F(S(k {+} N|k) {-} {S_d}).
\end{array}
\label{eq:23}
\end{aligned}
\end{align}

By substituting $\bar S(k)$ into the Eq.\Cref{eq:23}, the loss function is transformed as follows:
\begin{align}
\begin{array}{l}
J(k) = {(\bar S(k) - {{\bar S}_d})^T}\bar Q(\bar S(k) - {{\bar S}_d})\\
{\rm{          }} + \bar U{(k)^T}\bar R\bar U(k),
\end{array}
\label{eq:24}
\end{align}
where \(\bar S_d\) represents the reference state vector, \(\bar Q\) is a block diagonal matrix with the weighting matrix with a different weight \(F\) at the final step, and \(\bar R\) represents the weighting matrix associated with the input variables as follows:
\begin{align}
\bar {{S_d}}  = {\left( {{S_d},{S_d}, \cdots ,{S_d}} \right)^T},
\label{eq:25}
\end{align} 
\begin{align}
\bar Q  = diag\left( {Q,Q, \cdots ,Q,F} \right),
\label{eq:26}
\end{align} 
\begin{align}
\bar R  = diag\left( {R,R, \cdots ,R} \right).
\label{eq:27}
\end{align} 

Then, transform the loss $J(k)$ of Eq.\Cref{eq:24} into a form that is only related to $\bar U (k)$:
\begin{align}
\begin{array}{l}
J(k) {=} \left\| {\bar S (k) {-} \bar {{S_d}} } \right\|_{\bar Q }^2 {+} \left\| {\bar U (k)} \right\|_{\bar R }^2\\
 {=} \left\| {MS(k) {+} C(k)\bar U (k) {-} \bar {{S_d}} } \right\|_{\bar Q }^2 {+} \left\| {\bar U (k)} \right\|_{\bar R }^2.
\end{array}
\label{eq:28}
\end{align} 

Considering the feasible range of actual system inputs and state variables, the constraints imposed on the system are as follows:
\begin{align}
{U_{\min }} \le U(k) \le {U_{\max }},
\label{eq:29}
\end{align} 
\begin{align}
{S_{\min }} \le S(k) \le {S_{\max }},
\label{eq:30}
\end{align} 
where the parameters ${U_{\min }}$, ${U_{\max }}$, ${S_{\min }}$ and ${S_{\max }}$ denote the lower bounds and upper bounds of the input and state variables, respectively. Extending to an $N \times 2$ dimensional matrix form:
\begin{align}
\left\{ {\begin{array}{*{20}{l}}
{{{\bar U }_{\min }} = {{({U_{\min }},{U_{\min }}, \cdots {U_{\min }})}^T}}\\
{{{\bar U }_{\max }} = {{({U_{\max }},{U_{\max }}, \cdots {U_{\max }})}^T}}\\
{{{\bar S }_{\min }} = {{({S_{\min }},{S_{\min }}, \cdots {S_{\min }})}^T}}\\
{{{\bar S }_{\max }} = {{({S_{\max }},{S_{\max }}, \cdots {S_{\max }})}^T}}.
\end{array}} \right.
\label{eq:31}
\end{align} 

The constraints of the prediction interval are as follows:
\begin{align}
\left\{ {\begin{array}{*{20}{l}}
{{\rm{ }}{{\bar U }_{\min }} \le \bar U (k) \le {{\bar U }_{\max }}}\\
{{{\bar S }_{\min }} \le MS(k) + C(k)\bar U (k) \le {{\bar S }_{\max }}}
\end{array}} \right.{\rm{ }}.
\label{eq:32}
\end{align} 

Transform constraints into matrix form: 
\begin{align}
W = {\left( {\begin{array}{*{20}{c}}
{{I_{2N}}}&{ - {I_{2N}}}&{C(k)}&{ - C(k)}
\end{array}} \right)^T},
\label{eq:33}
\end{align} 
\begin{align}
w = {\left( {\begin{array}{*{20}{c}}
{{{\bar U }_{\max }}}\\
{ - {{\bar U }_{\min }}}\\
{{{\bar S }_{\max }} - MS(k)}\\
{ - {{\bar S }_{\min }} + MS(k)}
\end{array}} \right)^T},
\label{eq:34}
\end{align} 
then, the optimization problem can be written as follows:
\begin{align}
\left\{ \!\!\begin{array}{l}
\mathop {{\rm{min}}}\limits_{\bar U(k)} \left\| {MS(k) {+} C(k)\bar U(k) {-} \bar {{S_d}} } \right\|_{\bar Q}^2 {+} \left\| {\bar U(k)} \right\|_{\bar R}^2\\
{\rm{s}}.{\rm{t}}.W\bar U(k) \le w.
\end{array}\!\!\! \right.
\label{eq:35}
\end{align}  

Due to the significant property of convex optimization problems \citep{Bet2022}, in which a local optimal solution is also a global optimal solution, several efficient algorithms have been devised to address these problems. As a result, this paper transforms the optimization problem Eq.\Cref{eq:35} into a conventional convex quadratic programming format, as follows:
\begin{align}
\left\{ \begin{array}{l}
\mathop {{\rm{min}}}\limits_{\bar U(k)} \frac{{\rm{1}}}{{\rm{2}}}{{\bar U}^T}(k)H(k)\bar U(k) + {h^T}(k)\bar U(k)\\
{\rm{s}}.{\rm{t}}.{\rm{ }}W\bar U(k) \le w,
\end{array} \right.
\label{eq:36}
\end{align}  
where \(H(k)\) represents the second-order partial derivatives of the cost function with respect to the \(\bar U(k)\), \(h(k)\) represents the gradient vector of the cost function with respect to the \(\bar U(k)\), which can be calculated as follows:
\begin{align}
H(k) = 2({C^T}(k)\bar Q C(k) + \ R ),
\label{eq:37}
\end{align}  
\begin{align}
h(k) = {C^T}(k)\bar Q (MS(k) - \bar {{S_d}} ).
\label{eq:38}
\end{align}  
\section{Methodology}
\label{Section:3}
\subsection{Flexible Correction-Model Predictive Control}
\label{Section:3.1}
This section provides a detailed introduction to the Fc-MPC algorithm, which is based on flexible correction. This algorithm addresses the issue that existing model predictive control methods lack dynamic prediction and model updating, leading to inadequate anti-interference performance. 

Considering $\xi (k)$ as the unknown random interference existing in the actual navigation system at time \textit{k}. The model with interference is constructed as follows:
\begin{align}
S(k + 1) = AS(k) + B(k - 1)U(k) + \xi (k).
\label{eq:39}
\end{align} 

Combining Eq.\Cref{eq:13} and Eq.\Cref{eq:39} the unknown random interference is calculated as follows:
\begin{align}
\xi (k) = [B(k) - B(k - 1)]U(k).
\label{eq:40}
\end{align}

When introduce the ${U_a}(k)$ to compensate the input, the input of the model is:
\begin{align}
{U_h}(k) \buildrel \Delta \over = U(k) - {U_a}(k),
\label{eq:41}
\end{align}
where the compensated input variable is ${U_h}(k)$ and the state with unknown random interference is:
\begin{align}
S(k{+}1) {=} AS(k){+}B(k {-} 1)[{U_h}(k) {+} {U_a}(k)] {+} \xi (k).
\label{eq:42}
\end{align}

Then, the following equation is the conditions for uniform compensation:
\begin{align}
B(k - 1){U_a}(k) =  - \xi (k).
\label{eq:43}
\end{align} 

Finally, combining the Eq.\Cref{eq:41}, Eq.\Cref{eq:42} and Eq.\Cref{eq:43} the system state prediction model after compensation is:
\begin{align}
S(k + 1) = AS(k) + B(k - 1){U_h}(k).
\label{eq:44}
\end{align}

According to \Cref{Section:2.3}, it is converted into the standard convex quadratic optimization form:
\begin{align}
\left\{ \begin{array}{l}
\mathop {{\rm{min}}}\limits_{\bar U (k)} {\rm{  }}\frac{{\rm{1}}}{{\rm{2}}}{\rm{ }}\bar{U}_h^T(k)H(k)\bar {U}_h (k) + {h^T}(k)\bar {U}_h (k)\\
{\rm{s}}{\rm{.t}}{\rm{.   }}W\bar {{U_h}} (k) \le w.
\end{array} \right.
\label{eq:45}
\end{align}

\subsection{Fc-MPC for Combined Navigation Systems}
\label{Section:3.2}
\subsubsection{Combined navigation systems state equation}
\label{Section:3.2.1}
When there is a slight alteration in the spatial distribution of the local magnetic gradient, the precision of the heading angle computation diminishes \citep{Yu2023,Yang2022}.The proposed Fc-MPC algorithm uses a short term interval to combine the results of the inertial navigation error equation with the position error from geomagnetic navigation using a Kalman filter. This makes the geomagnetic navigation system more accurate and stable. The state equation for the combined geomagnetic and inertial navigation system is as follows:
\begin{align}
\left\{ \begin{array}{l}
\mathop \delta {X_{(k)}} = F\delta {X_{(k - 1)}} + {w_{(k - 1)}}\\
w \sim N{\rm{(0,}}~{Q_c}{\rm{)}},
\end{array} \right.
\label{eq:46}
\end{align}  
where ${w_{(k-1)}}$ denotes the process noise; \(F\) denotes the state matrix, which follows a normal distribution with an expected value of zero; ${Q_c}$ denotes the covariance matrix denoted.

The state error propagation equation of the combined system, which defines the state variable $\delta X $, is as follows:
 \begin{align}
\begin{array}{l}
\delta X = (\alpha ,\beta ,\gamma ,\delta \lambda ,\delta \varphi ,\delta h,\delta {V_x},\delta {V_y},\delta {V_z},\\
{\rm{         }}{\varepsilon _{cx}},{\varepsilon _{cy}},{\varepsilon _{cz}},{\varepsilon _{rx}},{\varepsilon _{ry}},{\varepsilon _{rz}}{)^T},
\end{array}
\label{eq:47}
\end{align}  
where $\alpha$ and $\gamma$ represent the horizontal error angle of the inertial system and the heading error angle, respectively. The variable $\beta$ represents the prior information distributed over $\delta X$. The terms $\delta\lambda$, $\delta \varphi$, and $\delta h$ represent the errors in longitude, latitude, and height, respectively. Additionally, $\delta {V_x}$, $\delta {V_y}$, and $\delta {V_z}$ represent the errors in eastward, northward, and vertical velocities, respectively. The ${\varepsilon _{cx}}$, ${\varepsilon _{cy}}$, and ${\varepsilon _{cz}}$ refer to the constant drift of the eastward, northward, and azimuth gyros, respectively. Meanwhile, ${\varepsilon _{rx}}$, ${\varepsilon _{ry}}$, and ${\varepsilon _{rz}}$ represent the random drifts of the eastward, northward, and azimuth gyros, respectively.

Since the inherent short-term stability of inertial systems in heading calculations can lead to cumulative errors during long-term navigation missions, only the state variables at the next sampling time of the inertial navigation system are utilized in areas characterized by low spatial variability of the geomagnetic gradient field.

\subsubsection{Combined navigation systems state update}
\label{Section:3.2.2}
The presence of various interferences, such as geomagnetic anomalies, geomagnetic storms, and inertial drift, during the navigation process can affect the accuracy and stability of heading calculations. Therefore, this paper constructs an observation model to estimate the magnitude of the unknown error. The following details outline how to implement this model for updating the state of combined navigation systems using Fc-MPC:

\noindent\textbf{1)~Observation model:} Assuming the inertial navigation measured location \(Z_m\) is as follows:
\begin{align}
{Z_m} = {({\lambda _m},{\varphi _m})^T}.
\label{eq:48}
\end{align}  
where \(\lambda_m\) and \(\varphi_m\) denote the longitude and latitude measured from the inertial system. The geomagnetic navigation calculates the location \(Z_c\), which is defined as follows:
\begin{align}
{Z_c} = {({\lambda _c},{\varphi _c})^T}.
\label{eq:49}
\end{align} 
where \(\lambda_c\) and \(\varphi_c\) denote the longitude and latitude calculated from the geomagnetic navigation system. The geomagnetic elements for the observation equation can be obtained from the WMM2020 \citep{NCEI2020} model or by real-time measurements with a triaxial magnetometer. 

\noindent\textbf{2) State observation equation:} Combining the state variable from Eq.\Cref{eq:42}. The observation equation of the system is as follows: 
\begin{align}
\left\{ \begin{array}{l}
\mathop \delta {Z_{(k)}} {=} {H_B}\delta {X_{(k {-} 1)}} {+} {v_{(k {-} 1)}} {=} {Z_{m(k)}} {-} {Z_{c(k)}}\\
v \sim N(0,{R_c}),
\end{array} \right.
\label{eq:50}
\end{align}  
where \(v_{(k-1)}\) denotes the measurement noise, which follows a normal distribution with an expected value of zero and a covariance matrix represented by \(R_c\); $\delta X$ denotes the state variable established by the prior state estimate \(\delta \hat X_{(k)}^ -\); ${H_B}$ is a \(2 \times 15\) dimensional observation matrix to extract only the longitude and latitude dimensions from the state of Eq.\Cref{eq:47}, which is calculated as follows:
\begin{align}
{H_B} = \left( {\begin{array}{*{20}{c}}
{{0_{2 \times 3}}}&{{I_{2 \times 2}}}&{{0_{2 \times 10}}}
\end{array}} \right),
\label{eq:51}
\end{align}
where \({0_{2 \times 3}}\) represents a \(2 \times 3\) zero matrix; \({{I_{2 \times 2}}}\) is a \(2 \times 2\) identity matrix; \({0_{2 \times 10}}\) represents a \(2 \times 10\) zero matrix.

\noindent\textbf{3) Kalman filter gain update:} To integrate the inertial and geomagnetic navigation data, the Kalman filter is employed. The error location \(\delta {Z_{(k)}}\) obtained by inertial navigation and geomagnetic navigation systems. The data fusion process is outlined as follows:

\begin{align}
\delta \hat X_{(k)}^ -  = F\delta {\hat X_{(k - 1)}},
\label{eq:52}
\end{align}  
where \(\delta \hat X_{(k)}\) denotes state estimate of $\delta X$. The prior state estimation covariance matrix \(P_{(k)}^ - \) is obtained as follows. 
\begin{align}
P_{(k)}^ -  = F{P_{(k - 1)}}{F^T} + {Q_c},
\label{eq:53}
\end{align}  
where \(P_{(k)}\) denotes error covariance matrix. The Kalman filter gain\(K_k\) update by ${H_B}$ is as follows: 
\begin{align}
{K_{(k)}} = \frac{{P_{(k)}^ - H_B^T}}{{{H_B}P_{(k)}^ - H_B^T + {R_c}}},
\label{eq:54}
\end{align} 

\noindent\textbf{4) Correction and feedback:} The state update process of geomagnetic/inertial combined navigation system is as follows:
\begin{align}
\delta {\hat X_{(k)}} = \delta \hat X_{(k)}^ -  + {K_{(k)}}(\delta {Z_{(k)}} - {H_B}\delta \hat X_{(k)}^ - ),
\label{eq:55}
\end{align}  
where $\delta {\hat X_{(k)}}$ is the posterior state estimate of the geomagnetic field resulting from data fusion between the measurement location error $\delta{Z_{(k)}}$ and the prior state estimate $\delta \hat X_{(k)}^ - $.

The updated posterior state estimation error covariance matrix is as follows:
\begin{align}
{P_{(k)}} = ({I_{15}} - {K_{(k)}}{H_B})P_{(k)}^ - .
\label{eq:56}
\end{align} 

The correction value $\delta {\hat X_{(k)}}$ is used as output feedback and added to the location of the geomagnetic navigation system as a correction term. 

\begin{align}
\delta {\hat X_{(k)}} = {\left( {\begin{array}{*{20}{c}}
{{0_{1 \times 3}}}&{\delta {{\hat \lambda }_{(k)}}}&{\delta {{\hat \varphi }_{(k)}}}&{{0_{10 \times 3}}}
\end{array}} \right)^T},
\label{eq:57}
\end{align} 
where \(\delta {{\hat \lambda }_{(k)}}\) denotes longitude correction term and \(\delta {{\hat \varphi }_{(k)}}\) denotes latitude correction term. 
\begin{align}
(\lambda_{c(k+1)},\!\varphi_{c(k+1)}){^T}{=}(\lambda_{c(k)},\!\varphi_{c(k)})^T\!{+}(\delta {{\hat \lambda }_{(k)}},\!\delta {{\hat \varphi }_{(k)}})^T\!\!.
\label{eq:58}
\end{align} 

The Fc-MPC algorithm for combined navigation systems in this paper is realized based on these modifications. 

\noindent\textbf{5) Determine terminate condition:} The optimization process iterates until the distance from the current position to the destination is below a predefined threshold \(\varepsilon\), which is set according to the actual navigation accuracy requirements.

The complete Fc-MPC for combined navigation system is summarized as \Cref{Algorithm:Fc-MPC Combined}.
\begin{algorithm}
\caption{Fc-MPC algorithm for geomagnetic and inertial combined navigation system.}
\label{Algorithm:Fc-MPC Combined}
\SetAlgoLined
\KwIn{${S}_0$,${S}_d$,$\lambda_0$,$\lambda_d$,$\varphi_0$,$\varphi_d$,$\theta_1$,$\theta_2$,$v_1$,$v_2$,$k$=2,\\
      \qquad Fc-MPC parameters: ${A}$,${B}$,$N$,${Q}$,${F}$,${R}$,\\
      \qquad \text{Data fusion} parameters:{$F$},${Q}_c$,${R}_c$,${P}_0$,${H}_B$}          \textbf{Constraints}:${U}_{min}$,${U}_{max}$,${S}_{min}$,${S}_{max}$, ${S}(0)$=${S}_0$   
      Calculate \(d\) from the current location to destination\;
   \While{$d$>$\epsilon$}
   {
        \eIf{$k$==2}
        {
        ${G}(k)$\ calculated by Eq.\Cref{eq:5}-Eq.\Cref{eq:9}\;
        }
        {
        Update ${G}(k+1)$ by Eq.\Cref{eq:10};
        }
        Solve the optimization problem by Eq.\Cref{eq:45}\;
        Take the first result of $U(k)$, $u^*(k)$\;
        Calculate $v_k$ and $\theta_k$ by Eq.\Cref{eq:3}-Eq.\Cref{eq:4}\;
        Update $S(k+1)$ by Eq.\Cref{eq:44}\;
        Calculate geomagnetic location \((\lambda_{c(k)},\!\varphi_{c(k)})^T\) ;\\     
        \If{min({G}($k$))<$\sigma$}
        {
        Initializing $\delta {X}_k$ by Eq.\Cref{eq:46}\;
Get inertial system location ($\lambda_{m(k)}$,$\varphi_{m(k)}$)\(^T\)\;
        Calculate $\delta {Z}_k$ by Eq.\Cref{eq:50}\;
        Calculate $\delta\hat{{X}}_{k}$ by Eq.\Cref{eq:52}-Eq.\Cref{eq:55};\\
        Calculate correction location ($\delta {{\hat \lambda }_{(k)}}$,$\delta {{\hat \varphi }_{(k)}}$)\(^T\) by Eq.\Cref{eq:55}-Eq.\Cref{eq:57}\;
        Update geomagnetic system location $(\lambda_{c(k+1)}$, $\varphi_{c(k+1)})$\(^T\) and inertial system location $(\lambda_{m(k+1)}$,$\varphi_{m(k+1)})$\(^T\)\;
        Update $(\lambda_{k+1}$,$\varphi_{k+1}$)\(^T\) by Eq.\Cref{eq:58}\;}
        Update iteration time $k$=$k$+1, and distance $d$;\\ 
  }
\KwOut{Location ($\lambda$,$\varphi$)\(^T\)}
\end{algorithm}

%%%%%%%%%%%%%%%%%%%%%%%%%%%%%%%%%%%%%%%%%%%%%%%%%%%%%%%%%%%
\section{Simulation and Experiment}
\label{Section:4}
The simulation platform is as follows: CPU: AMD Ryzen 7 5800H; Memory: 16.00GB; Main frequency: 3.20GHz; Software environment: MATLAB 2023b. The simulated experimental area selected is a rectangular region in the Pacific Ocean, which extends from (145\(^\circ\)E, 27\(^\circ\)N) to (165\(^\circ\)E, 34\(^\circ\)N). This study designates the coordinates (152\(^\circ\)E, 33\(^\circ\)N) as the starting location and (158\(^\circ\)E, 28\(^\circ\)N) as the destination location. In all scenarios, \textit{D} and \textit{I} real-time information are obtained from WMM2020 model. 

\begin{table}[ht]
\centering
\caption{PSINS inertial navigation simulation error settings.}
\label{tab:1}
\fontsize{9}{9}\selectfont
\resizebox{\columnwidth}{!}{%
\begin{tabularx}{\columnwidth}{X l}
\toprule
\multicolumn{1}{l}{\textbf{Parameter description}} & \multicolumn{1}{l}{\textbf{Value}} \\ \midrule
Platform misalignment
angle (arcmin) & [50;50;500] \\
Speed error (m/s) & 10 \\
East position error (m) & 5000 \\
North position error (m) & 5000 \\
Vertical position error (m) & 0 \\ \bottomrule
\end{tabularx}%
}
\end{table}

\begin{table}[ht]
\centering
\caption{Inertial navigation simulation parameters settings.}
\label{tab:2}
\fontsize{9}{9}\selectfont
%\resizebox{\columnwidth}{!}
{%
\begin{tabularx}{\columnwidth}{X l l}
\toprule
\textbf{Parameters description}             & \textbf{Value}           \\ \midrule
Yaw angle (\(^\circ\))           & 2.3123                             \\
Speed $v$ (m/s)                  & 20.5778                           \\
Eastward speed $v_e$ (m/s)        & 15.1756                            \\
Northward speed $v_n$ (m/s)       & -13.8977                           \\
Vertical speed $v_u$ (m/s)        & 0                                  \\
Step length (s)                & 0.005                             \\
Sampling time (s)              & 0.01                              \\
Longitude (E\(^\circ\))     & 33                                \\
Latitude  (N\(^\circ\))     & 152                               \\
Altitude (km)              & 0                                  \\ \bottomrule
\end{tabularx}%
}
\end{table}

\begin{table}[ht]
\centering
\caption{Data fusion related parameter settings.}
\label{tab:3}
\fontsize{9}{9}\selectfont
{
\begin{tabularx}{\columnwidth}{X l l}
\toprule
\textbf{Parameters description}       & \textbf{symbol}             & \textbf{Value}                           \\ \midrule
Process noise matrix                 & {$Q_c$} & {{diag($0_{1\times3}$,0.05,0.05,$0_{1\times10}$)}} \\
Measurement noise matrix             & {$R_c$} & {{diag(2,2)}}       \\
Posteriori matrix & $P_0$                        & {diag($0_{1\times3}$,1,1,$0_{1\times10}$)}                              \\
State matrix                         & {$F$}                           & {$I_{15}$}                              \\
Measurement matrix                   & $H_B$                        &{($0_{2\times3}$,$I_{2}$,$0_{2\times10}$)}                              \\ \bottomrule
\end{tabularx}%
}
\end{table}

In this paper, the parameters relating to the absolute positioning error of inertial navigation are presented in \Cref{tab:1}. These parameters include platform misalignment angles, speed error, east position error, north position error, and vertical position error. The parameters for the inertial navigation simulation are detailed in \Cref{tab:2}, which encompass the cover yaw angle, speed, eastward speed, northward speed, vertical speed, step length, sampling time, longitude, latitude, and altitude. Data fusion parameters are outlined in \Cref{tab:3}, including the process noise matrix, measurement noise matrix, posteriori matrix, state matrix, input matrix, and measurement matrix. The parameters related to the Fc-MPC algorithm are provided in \Cref{tab:4}. The inertial navigation simulation employs a high-precision INS trajectory generator from the MATLAB PSINS toolbox \citep{Min2024}.

To assess the effectiveness of the proposed Fc-MPC algorithm, this section conducts a comparative simulation analysis with linear time-invariant (LTI-MPC) \citep{Nguyen2023} and linear time-variant (LTV-MPC) \citep{Mousavi2013} algorithms. LTI-MPC is applicable to systems with constant parameters, whereas LTV-MPC is suited for systems with time-varying parameters. This paper evaluates the performance of Fc-MPC with non-interference, magnetic storm, and real data experiments utilizing quantitative metrics to emphasize its advantages and identify areas for potential improvement. The specific process of simulation and experiment is as follows.

\begin{table}[htp]
\centering
\caption{Fc-MPC algorithm related parameters settings.}
\label{tab:4}
\fontsize{9}{9}\selectfont
{%
\begin{tabularx}{\columnwidth}{>{\raggedright\arraybackslash}X l l}
\toprule
\textbf{Parameter Description}  & \textbf{Symbol} & \textbf{Value}                  \\ \midrule
\textbf{Constraints} & & \\
\hspace{2pt}Minimum state         & $S_{min}$      & [-10; 10]                \\
\hspace{2pt}Maximum state         & $S_{max}$      & [100; 100]              \\
\hspace{2pt}Minimum control       & $U_{min}$      & [0; 0]                    \\
\hspace{2pt}Maximum control       & $U_{max}$      & [40; 40]                  \\ \midrule
\textbf{Mission Settings} & & \\
\hspace{2pt}Starting location        & $(\lambda_{o}, \varphi_{o})$\(^T\)        & (152\(^\circ\)E, 33\(^\circ\)N)\(^T\)            \\
\hspace{2pt}Destination location          & $(\lambda_{d}, \varphi_{d})$\(^T\)          & (158\(^\circ\)E, 28\(^\circ\)N)\(^T\)            \\ \midrule
\textbf{Algorithm Parameters} & & \\
\hspace{2pt}Prediction Interval   & $N$            & 2                             \\
\hspace{2pt}Sampling period       & $T$            & 10                            \\
\hspace{2pt}Heading angle         & ($\theta_1, \theta_2$)     & (0, 270)                             \\
                                   \hspace{2pt}Speed (km/h)          & $v$            & 50                            \\ \midrule
\textbf{Matrices and Weights} & & \\
\hspace{2pt}System matrix         & $A$            & [1 0; 0 1]              \\
\hspace{2pt}Input matrix          & $B$            & $G \cdot T$                  \\
\hspace{2pt}Process noise matrix    & $Q$            & [1 0; 0 1]               \\
\hspace{2pt}State transition matrix & $F$          & [1 0; 0 1]               \\
\hspace{2pt}Control weight matrix & $R$            & [10 0; 0 10]             \\ 
\bottomrule
\end{tabularx}%
}
\end{table}

\subsection{Simulation without Interference }
\label{Section:4.1}
The primary purpose of testing different algorithms in non-interference simulations is to establish a baseline for their performance under ideal conditions. This approach aids in understanding the inherent capabilities and limitations of each algorithm, including convergence rate, precision, and trajectory efficiency without the influence of external disturbances, and the results presented in \Cref{tab:5}.

\begin{table}[htb]
\centering
\caption{Navigation approaches performance analysis.}
\label{tab:5}
\fontsize{9}{9}\selectfont
\begin{tabularx}{\columnwidth}{l X X X}
\toprule
\textbf{Metric} & \textbf{LTI-MPC} & \textbf{LTV-MPC} & \textbf{Fc-~MPC} \\ \midrule
\multicolumn{4}{l}{\textbf{Accuracy Performance}} \\
\hspace{2pt}CEP (km) & 35.18      & 8.92       & 0.19      \\
\hspace{2pt}Terminal longitude (E\(^\circ\))  & 157.68     & 157.92     & 158.01       \\
\hspace{2pt}Terminal latitude (N\(^\circ\)) & 28.15      & 28.04      & 28.02        \\

\midrule
\multicolumn{4}{l}{\textbf{Stability Performance}} \\
\hspace{2pt}Mean deviation (km) & 0.76       & 0.75       & 0.25      \\
\hspace{2pt}Max deviation (km)  & 1.41       & 1.42       & 0.35      \\
\hspace{2pt}PMR (\%)  & 26.04      & 26.32      & 90.85     \\ 

\midrule
\multicolumn{4}{l}{\textbf{Efficiency Performance}} \\
\hspace{2pt}Iterations              & 171        & 169        & 164       \\
\hspace{2pt}Variability (km\(^2\)) & 0.19       & 0.18       & 0.15      \\
\hspace{2pt}Trajectory length (km)        & 864.59     & 837.52     & 812.11    \\ 
\bottomrule
\end{tabularx}
\end{table}

\subsubsection{Accuracy evaluate of the approaches}
\label{Section:4.1.1}
The remaining distance from the navigation terminal location to the destination was calculated to assess the accuracy of the results using the following metrics: circular error probable and average terminal location, determined through Monte Carlo simulation \citep{Madadi2020}. The navigation results are listed in \mbox{\Cref{tab:5}}. These metrics are defined as follows:

\begin{itemize}[leftmargin=*]
    \item \textbf{The circular error probable (CEP):} The distance between the final position of the navigation and the destination when the iteration stops. The CEP denotes the circular error probable \citep{Zhang2012}, and it is calculated as follows:
    \begin{equation}
\text{CEP} {=} \sqrt{\frac{1}{N} \sum_{i=1}^{N} ((x_i {-} x_d)^2 {+} (y_i {-} y_d)^2) \cdot I(d_i \leq d_{50})},
\label{eq:59}
\end{equation}
where \(N\) denotes the total number of tests conducted. The coordinates \((x_i, y_i)\) specify the last arrival position of the \(i\)-th Monte Carlo simulation; \(d_{50}\) denotes the median distance, i.e., the probability that at least 50\% of the ending locations fall within some circular area centered on the destination location; \(I\) denotes the indicator function, which takes the value of \(1\) when \({d_i \le {d_{50}}}\).
    \item \textbf{Terminal Location:} The \(N=50\) times simulation average latitude and longitude when the approaches terminal are considered the terminal location result.
\end{itemize}

\begin{figure}[htp]
    \centering
    \includegraphics[width=0.7\linewidth]{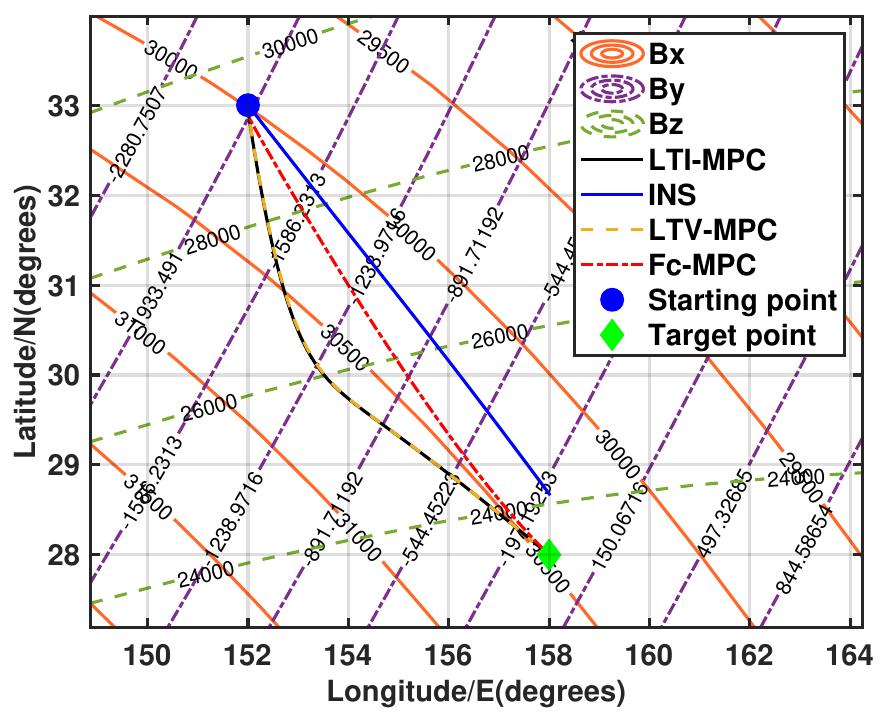}
    \caption{Simulated navigation trajectory without interference.}
    \vspace{-5pt}
    \label{fig:4}
\end{figure}

The accuracy performance shown in \Cref{tab:5} outcomes supports the superiority of the Fc-MPC algorithm. The CEP result of the LTI-MPC algorithm is 35.18 km, and the LTV-MPC algorithm is 8.92 km. In comparison, the Fc-MPC algorithm achieves the CEP to just 0.19 km. The average terminal location result of the Fc-MPC algorithm is 2.25 km from the destination. This represents a reduction of 36.6\% compared to the LTI-MPC algorithm distance of 3.55 km and a reduction of 75.1\% compared to the LTV-MPC algorithm distance of 9.03 km. This demonstrates the ability to provide more accurate terminal location of the Fc-MPC algorithm. This advantage can also be seen in \mbox{\Cref{fig:4}}, the navigation trajectories obtained by the LTI-MPC and LTV-MPC algorithms are more twisted, but the Fc-MPC algorithm result is more rational.

\vspace{-5pt}

\subsubsection{Stability evaluate of the approaches}
\label{Section:4.1.2}
The stability evaluation of optimal trajectories in 50 Monte Carlo simulations of different MPC algorithms using three metrics: the mean path deviation, maximum path deviation, and path matching rate. The results are detailed in \Cref{tab:5}. These metrics have the following definitions:

\begin{itemize}[leftmargin=*]
    \item \textbf{Mean path deviation:} The mean of the distance differences between each position on the navigation route and the shortest trajectory is determined by the equation:
\begin{align}
\text{Dev}_{mean} {=} \frac{1}{n}\sum\nolimits_{i = 1}^n {\left| {{d_i} {-} {{\hat d}_i}} \right|}, 
\label{eq:60}
\end{align}
where \textit{n} is the number of iterations, ${d_i}$ is the position of the \textit{i}-th iteration, and ${\hat d_i}$ is the position of the \textit{i}-th step of the shortest trajectory.

\item \textbf{Maximum path deviation:} The distance of the farthest position on the navigation route from the shortest trajectory. The equation is as follows:
\begin{align}
\text{Dev}_{max} {=} \max \left( {\left| {{d_i}{-} {{\hat d}_i}} \right|} \right),
\label{eq:61}
\end{align}
where \textit{n} is the iterations, ${d_i}$ is the position of the \textit{i}-th iteration, and ${\hat d_i}$ is the position of the \textit{i}-th step of the shortest trajectory.

\item \textbf{Path matching rate:} The proportion of points on the navigation trajectory that match the shortest trajectory. The equation is as follows:
\begin{align}
\text{PMR}= \frac{{{p_{m}}}}{{{p_{a}}}} \times 100\%, 
\label{eq:62}
\end{align}
where ${p_{m}}$ is the number of points on the navigation path that match the optimal path, and ${p_{a}}$ denotes the aggregate of points along the shortest trajectory.
\end{itemize}

The results presented in \Cref{tab:5} illustrate the comparative performance of the LTI-MPC, LTV-MPC, and Fc-MPC algorithms in terms of navigation path accuracy. The LTI-MPC algorithm yields a mean path deviation of 0.76 km and a maximum path deviation of 1.41 km, along with a path matching rate (PMR) of only 26.04\%. In contrast, the LTV-MPC algorithm shows slightly improved results, with a mean path deviation of 0.75 km, a maximum path deviation of 1.42 km, and a PMR of 26.32\%. In contrast, the proposed Fc-MPC algorithm, however, demonstrates significant improvements over both the LTI-MPC and LTV-MPC algorithms. It achieves a mean path deviation of 0.25 km and a maximum path deviation of 0.35 km, reflecting a more accurate and consistent navigation trajectory. Furthermore, the Fc-MPC algorithm achieves a PMR of 90.85\%, indicating a much closer alignment with the optimal path compared to the other algorithms.

The improvements achieved by the Fc-MPC algorithm are substantial: the mean path deviation is reduced by 67.11\% compared to LTI-MPC and by 66.67\% compared to LTV-MPC. Similarly, the maximum path deviation decreases by 75.18\% and 75.35\%, respectively. The most notable enhancement is observed in the PMR, where the Fc-MPC algorithm shows an performance of 248.75\% over LTI-MPC and 245.23\% over LTV-MPC. The Fc-MPC algorithm in improving navigation path accuracy and stability compared to the other algorithms.

\subsubsection{Efficiency evaluate of the approaches}
\label{Section:4.1.3}
The average statistical results from 50 Monte Carlo simulations of various MPC algorithms are evaluated using three metrics: the number of iterations, variability, and trajectory length. These metrics are defined as follows:
\begin{itemize}[leftmargin=*]
    \item \textbf{Iterations:} The number of iterations required to find a path that meets the conditions.

    \item \textbf{Variability:} The path variability in step length reflects the degree of twist in the navigation path by measuring the change in distance at each step. The specific formula is as follows:
\begin{align}
\textit{Var} = \frac{1}{{n - 1}}\sum\nolimits_{i = 1}^n {{{({d_i} - \bar s)}^2}},
\label{eq:63}
\end{align}
where \textit{n} denotes the iterations; ${d_i}$ denotes the distance between the location of the \textit{i}-th step and the previous step; $\bar s$ is the mean step length.
    \item \textbf{Trajectory length:} According to the path generated by the navigation methods, the length of each section on the path is accumulated.
\end{itemize}

The results presented in \Cref{tab:5} highlight the superior performance of the Fc-MPC algorithm compared to the LTI-MPC and LTV-MPC algorithms. The Fc-MPC algorithm converges in just 164 iterations, which is fewer than the 171 iterations required by the LTI-MPC and the 169 iterations needed by the LTV-MPC. This demonstrates a more efficient computational process.

In terms of the step variance, which measures the stability of the trajectory, shows significant improvements with the Fc-MPC algorithm. The step variance for Fc-MPC is 0.15 km², representing a reduction of 21.05\% compared to the 0.19 km² of the LTI-MPC and 16.67\% lower than the 0.18 km² of the LTV-MPC. This reduction in variance indicates that the Fc-MPC algorithm produces more stable and smoother navigation paths.

Furthermore, for the trajectory length, the Fc-MPC algorithm achieves a total path length of 812.11 km, which represents a reduction of 6.07\% compared to the LTI-MPC's 864.59 km and a 3.03\% decrease compared to the LTV-MPC's 837.52 km. This indicates that the Fc-MPC method is more effective at minimizing path length, resulting in more optimized routes.

Overall, the Fc-MPC algorithm not only improves computational efficiency but also generates more optimized and stable trajectories, as illustrated in \Cref{fig:4}. This highlights the effectiveness of the Fc-MPC method compared to the other two algorithms in navigation applications.

\subsection{Simulation with Magnetic Storm }
\label{Section:4.2}
The primary goal of simulating navigation algorithms under magnetic storm conditions is to evaluate their robustness and effectiveness in the presence of significant geomagnetic disturbances. These simulations are crucial for understanding how well the algorithms can maintain trajectory accuracy, stability, and efficiency in real-world scenarios where environmental interference is common. Unlike non-interference simulations, which establish baseline performance under ideal conditions, magnetic storm simulations provide insights into each algorithm's adaptability and resilience to dynamic geomagnetic changes.
\subsubsection{Magnetic storm datasets for simulations}
\label{Section:4.2.1}
Magnetic storms, caused by solar activity such as solar wind and sunspots, lead to significant changes in the Earth's geomagnetic field. These storms typically progress through three phases: the initial, main, and recovery phases, lasting from several hours to a few days, during which geomagnetic components fluctuate significantly \citep{Li2024}. This study uses real geomagnetic data from global magnetic observatories, interpolated with Kriging algorithms to fill missing points \citep{Hu2024}. The dataset is divided into latitude blocks of 10°, and station data is categorized accordingly using the SuperMAG open-source code \citep{Gjerloev2024}. In this section, the simulation extracting the varying magnetic field data first performs Kriging interpolation for missing points before calculating total magnetic field strength d${B_H}$. The sign of $d{B_H}$ is based on the positive or negative sign of $d{B_x}$. Then, a table of global geomagnetic anomaly mapping is created based on latitude, longitude, $d{B_H}$, $d{B_x}$, and $d{B_y}$. The total magnetic field strength calculation equation is as follows:
\begin{align}
d{B_H} = \frac{{\left| {d{B_x}} \right|}}{{d{B_x}}} \times \sqrt {d{B_x}^2 + d{B_y}^2},
\label{eq:64}
\end{align}

\begin{figure}[htp]
    \centering
    \subfigure[]
    {
        \includegraphics[width=1\linewidth]{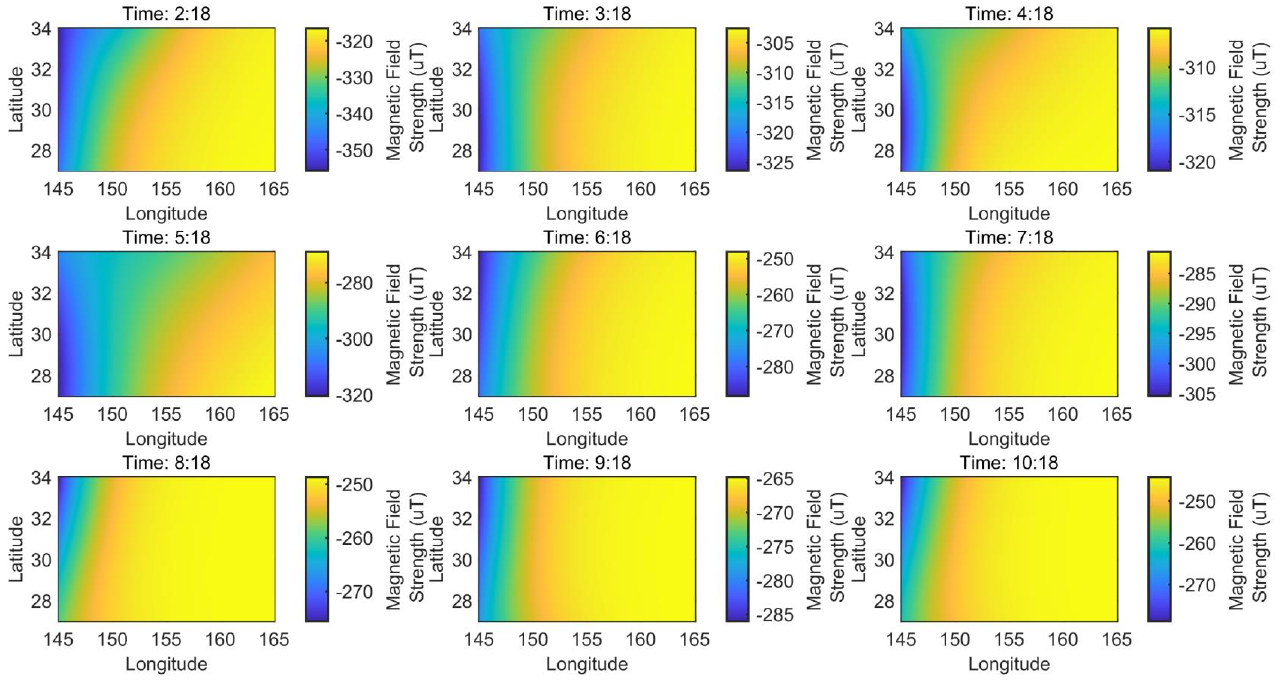}
        \label{fig:5a}
    }
    \subfigure[]
    {
        \includegraphics[width=0.95\linewidth]{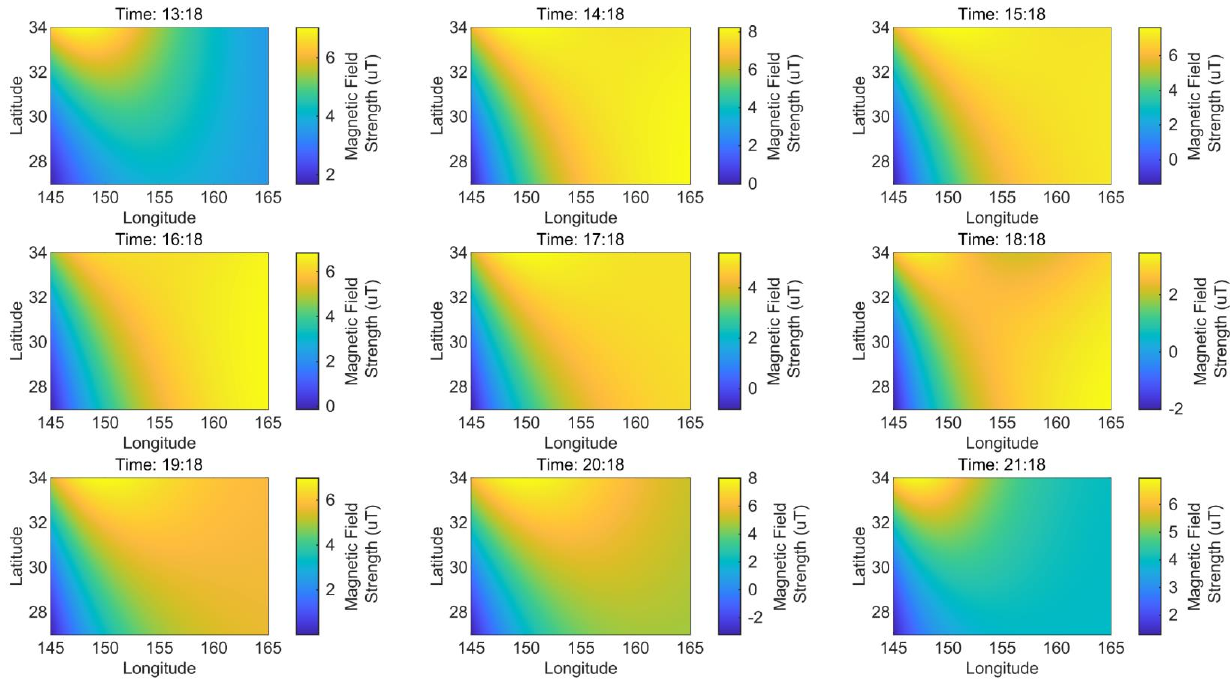}
        \label{fig:5b}
    }
    \caption{The datasets track the gradient of magnetic storm intensity every hour. (a) Long-term magnetic storm interference; (b) Short-term magnetic storm interference.}
    \label{fig:5}
\end{figure}

The solar flare released high-energy particles that increased the particle density and energy in the Earth's magnetosphere, disrupting the magnetic field and causing a geomagnetic storm \citep{DeMichelis2010}. To analyze the impact of different types of magnetic storm interference on the geomagnetic field. This paper selects two distinct periods for this analysis: 
\begin{itemize}[leftmargin=*]
    \item \textbf{Dataset for long-term, slowly varying interference:} This dataset covers a period from 2:00 AM to 8:00 AM Beijing Time on May 11, 2024. It captures a prolonged geomagnetic disturbance caused by active region AR3664, noted as the most severe geomagnetic storm since November 2004. Multiple full-halo coronal mass ejections occurred between May 8 and May 9, 2024, driving the storm \citep{Cun2024a}. This dataset, as shown in \Cref{fig:5a} is suitable for simulating and analyzing the effects of sustained magnetic fluctuations on navigation accuracy and stability over a longer period.

    \item \textbf{Dataset for short-term, high-intensity interference:} This dataset covers the period from 13:13 to 14:46 Beijing Time on May 6, 2024, during which an X4.5 X-ray flare erupted from active region AR3663. The flare caused a sharp increase in the particle density and energy in the Earth's magnetosphere, leading to rapid geomagnetic field changes that peaked at an orange alert level before subsiding at 14:46 \citep {Cun2024b}. This dataset, as shown in \Cref{fig:5b} is useful for evaluating the impact of intense solar activity on the performance of navigation algorithms.
\end{itemize}

\vspace{-10pt}
\subsubsection{Performance evaluation of navigation methods with long-term magnetic storm interference}
\label{Section:4.2.2}
\begin{figure}[ht]
    \centering
    \subfigure[]
    {       \includegraphics[width=0.45\linewidth]{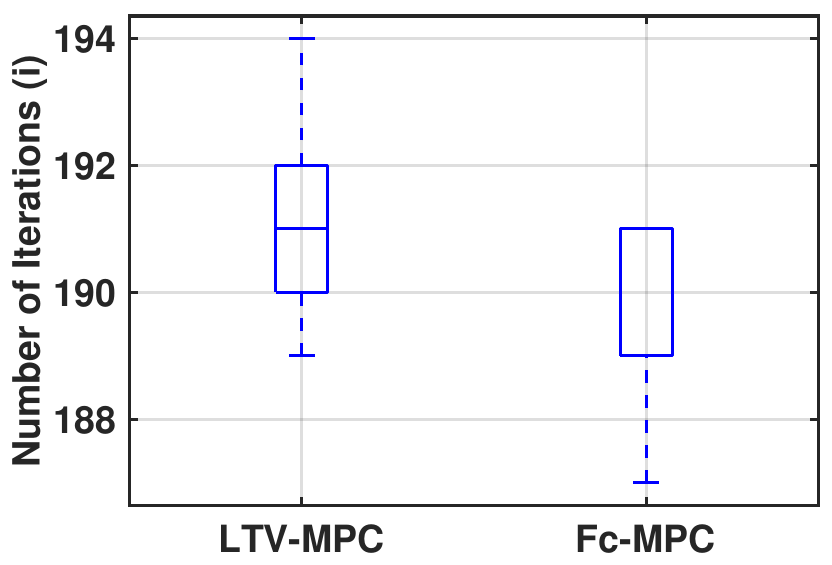}
        \label{fig:6a}
    }
    \subfigure[]
    {       \includegraphics[width=0.45\linewidth]{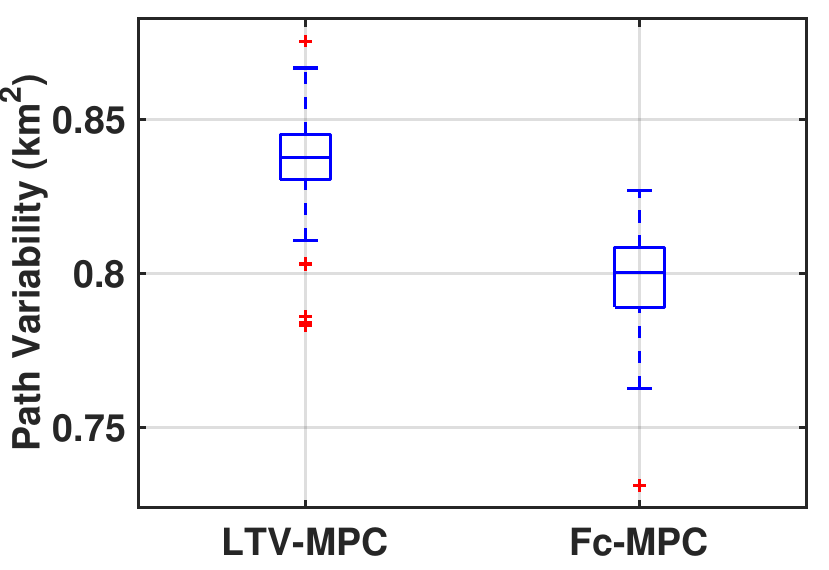}
        \label{fig:6b}
    }
        \subfigure[]
    {       \includegraphics[width=0.45\linewidth]{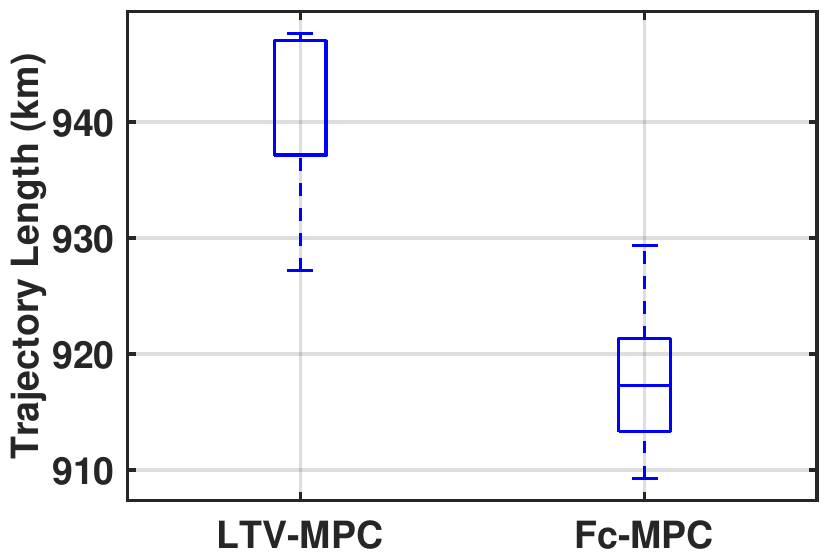}
        \label{fig:6c}
    }
    \subfigure[]
    {
        \includegraphics[width=0.45\linewidth]{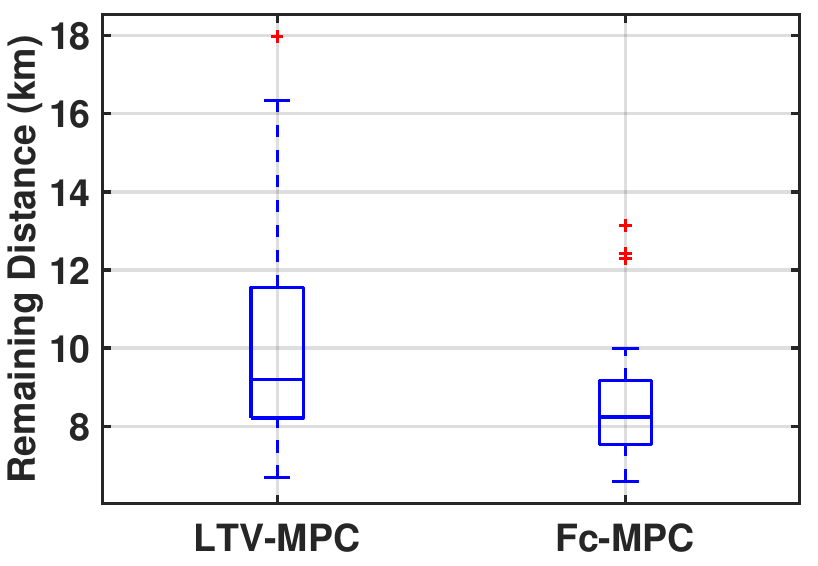}
        \label{fig:6d}
    }
    \caption{Statistical results of Monte Carlo simulations under long-term magnetic storm interference. (a) Number of iterations; (b) Path Variability; (c) Trajectory length; (d) Remaining distance to the destination.}
    \label{fig:6}
\end{figure}

\begin{figure}[htb]
    \centering \includegraphics[width=0.7\linewidth]{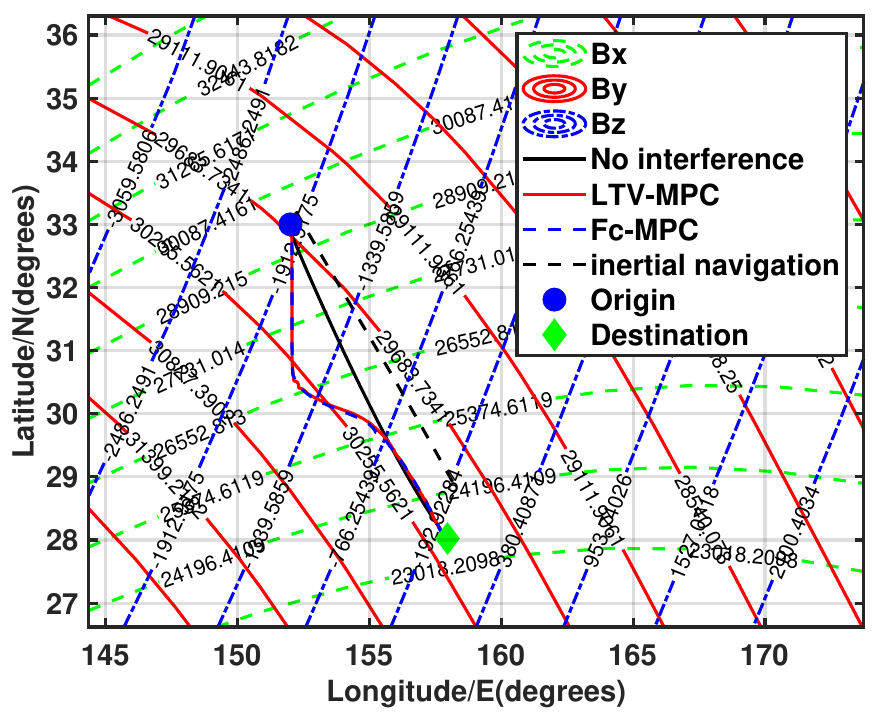}
    \caption{Simulated navigation optimal trajectories under long-term magnetic storm interference.}
    \label{fig:7}
\end{figure}

The results of constructing a long-term magnetic storm interference, as discussed in \Cref{Section:4.2.1}, for a simulated experimental area ranging from (27\(^\circ\)N, 145\(^\circ\)E) to (34\(^\circ\)N, 165\(^\circ\)E) are utilized to evaluate the navigation method in this section. The intensity of the magnetic storm varies over navigation time and is superimposed on the original WMM model. The LTI-MPC method is unsuitable for handling scenarios involving magnetic storms, as these storms introduce highly dynamic and nonlinear disturbances. Therefore, in this section, only the LTV-MPC and Fc-MPC methods are compared. The results from the Monte Carlo simulations were also collected to evaluate the stability of the methods, as shown in \Cref{fig:6}. Additionally, the optimal trajectories based on 50 Monte Carlo simulation results are presented in \Cref{fig:7}, while detailed performance metrics are shown in \Cref{tab:6}. 

\begin{table}[htb]
\centering

\fontsize{9}{9}\selectfont
\caption{Optimal performance comparison of LTV-MPC and Fc-MPC under long-term magnetic storm interference.}
\begin{tabularx}{\columnwidth}{lXX}
\toprule
\textbf{Metric} & \textbf{LTV-MPC} & \textbf{Fc-MPC} \\ \midrule
Signal-to-noise ratio (dB) & 5.09 & 5.25 \\
Iterations & 189 & 190 \\
Step Var (km\(^2\)) & 0.80 & 0.80 \\
Trajectory length (km) & 937.18 & 942.12 \\
Mean path deviation (km) & 0.60 & 0.59 \\
Max path deviation (km) & 1.59 & 1.56 \\
Remaining distance (km) & 9.60 & 7.60 \\
\bottomrule
\end{tabularx}%
\label{tab:6}
\end{table}

As shown in \Cref{tab:6}, it is evident that during long-term magnetic storms, the SNR of the Fc-MPC algorithm increases by 3.14\% compared to LTV-MPC, highlighting a slight improvement in noise resilience. The maximum path deviation in Fc-MPC also decreases by 1.89\%, indicating a more stable trajectory. Moreover, while the number of iterations in both algorithms is almost identical, the remaining distance for Fc-MPC is reduced by 20.83\% (2.00 km shorter), suggesting higher accuracy in reaching the destination. The trajectory length of the Fc-MPC method is 2.42\% shorter than that of the LTV-MPC method, indicating that LTV-MPC uses a more complex route to minimize error. This result can also be seen intuitively from \Cref{fig:7}.

The performance of the LTV-MPC algorithm and the Fc-MPC algorithm during long-term magnetic storms is illustrated in \Cref{fig:6}. This figure evaluates the results of 50 Monte Carlo simulations using several performance metrics, including the number of iterations, path variability, trajectory length, and remaining distance.

The performance of LTV-MPC and the Fc-MPC algorithm during long-term magnetic storms is shown in \Cref{fig:6a}. In comparison to the scenario without interference, the number of iterations increases significantly. However, Fc-MPC still demonstrates a faster convergence speed than the LTV-MPC method. Additionally, the range of the maximum (191) and minimum (187) values for the Fc-MPC iteration number is narrower than that of the LTV-MPC, which has a maximum of 194 and a minimum of 189. This observation highlights the stability of the Fc-MPC method. As shown in \Cref{fig:6b}, the path variability results from Monte Carlo simulations under long-term magnetic storm interference indicate that the LTV-MPC method exhibits significant fluctuations, ranging from 0.78 km$^2$ to 0.88 km$^2$, and includes six outliers, which indicates a low level of stability. In contrast, the variance distribution of the Fc-MPC method is more concentrated, spanning from 0.73 km$^2$ to 0.83 km$^2$, which shows that the Fc-MPC method achieves minimal path variability and greater stability.

The trajectory length of the navigation path generated by the LTV-MPC algorithm, as shown in the \Cref{fig:6c}, exhibits a high degree of stability and concentration; however, it is generally longer, with a median length of 937.15 km and a maximum value approaching 947.62 km. This suggests that, in certain instances, a longer path may be produced. In contrast, the Fc-MPC algorithm demonstrates a shorter path length, with a median of 917.29 km, and its maximum value of 929.35 km is significantly lower than the maximum value of the LTV-MPC algorithm (947.62 km). This indicates that even in a magnetic storm environment, the trajectory planned by the Fc-MPC algorithm remains efficient. Finally, \Cref{fig:6d} shows that the LTV-MPC algorithm exhibits a significant range of fluctuations in the distance from the actual arrival position to the intended destination, spanning from 6.68 km to 17.98 km. This algorithm also demonstrates a high median distance of 9.19 km and the presence of outliers, which may adversely impact overall positioning accuracy. In contrast, the distance distribution associated with the Fc-MPC algorithm is more concentrated, ranging from 6.59 km to 13.14 km, and features a lower median of 8.23 km, indicating better positioning accuracy.

In conclusion, the Fc-MPC method consistently outperforms the LTV-MPC method in terms of convergence speed, path variability, trajectory length, and positioning accuracy under long-term magnetic storm interference. This makes it a more stable and efficient solution for combined navigation systems in such environments, a notion further substantiated by the navigation trajectory shown in \Cref{fig:7}.

\subsubsection{Performance evaluation of navigation methods with short-term magnetic storm interference}
\label{Section:4.2.3}

\begin{figure}[h]
    \centering
    \subfigure[]
    {
        \includegraphics[width=0.45\linewidth]{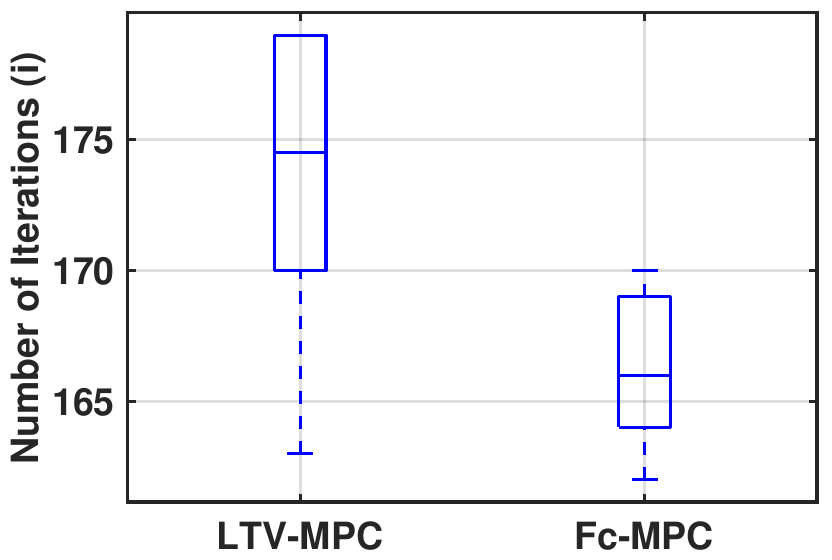}
        \label{fig:8a}
    }
    \subfigure[]
    {
        \includegraphics[width=0.45\linewidth]{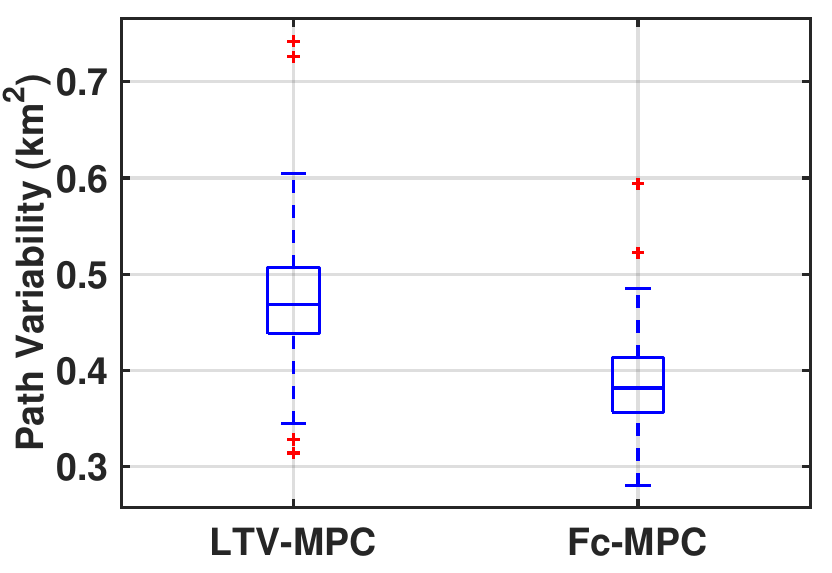}
        \label{fig:8b}
    }
        \subfigure[]
    {
        \includegraphics[width=0.45\linewidth]{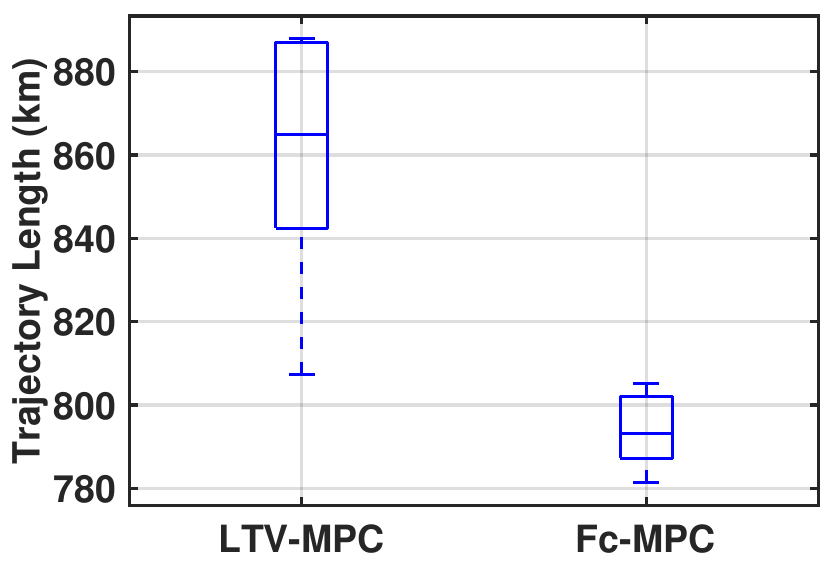}
        \label{fig:8c}
    }
    \subfigure[]
    {
        \includegraphics[width=0.45\linewidth]{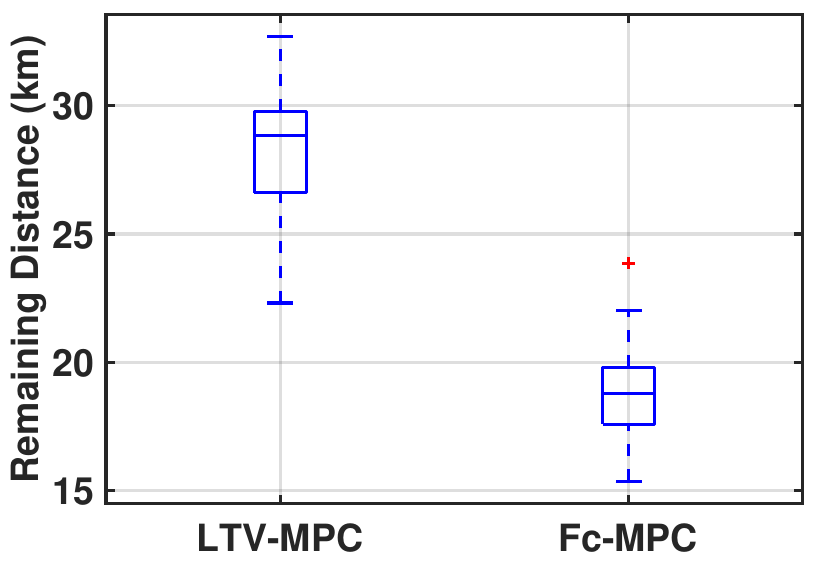}
        \label{fig:8d}
    }
    \caption{Statistical results of Monte Carlo simulations under short-term magnetic storm interference. (a) Number of iterations; (b) Path Variability; (c) Trajectory length; (d) Remaining distance to the destination.}
    \label{fig:8}
\end{figure}

\begin{figure}[htb]
    \centering \includegraphics[width=0.7\linewidth]{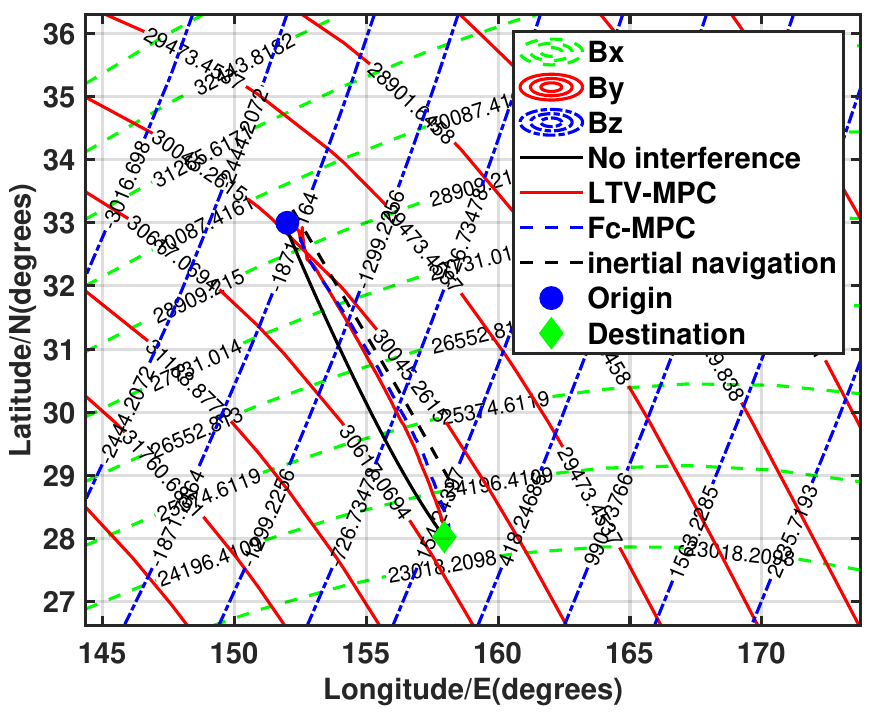}
    \caption{Simulated navigation optimal trajectories under short-term magnetic storm interference.}
    \label{fig:9}
\end{figure}

Due to the interference caused by long-term magnetic storms, we have evaluated the stability and accuracy of the methods discussed in \Cref{Section:4.2.2}, with a particular focus on the impact of accumulated errors. To further evaluate the performance of the Fc-MPC and LTV-MPC methods under the influence of short-term, high-intensity magnetic storm interference, this section utilizes the short-term magnetic storm datasets presented in \Cref{Section:4.2.1}. These datasets are superimposed on the WMM model to complete this section of the simulation, and the simulation area remains the same as that described in \Cref{Section:4.2.2}. The results from the 50 Monte Carlo simulations were also collected to evaluate the stability of the methods, as shown in \Cref{fig:8}. Additionally, the optimal trajectories based on 50 Monte Carlo simulation results are presented in \Cref{fig:9}, while detailed performance metrics are shown in \Cref{tab:7}.

In the case of short-term magnetic storms, the optimal performance shown in \Cref{tab:7} indicates that the trajectory length of the Fc-MPC algorithm (827.00 km) is reduced by 20.53 km compared to LTV-MPC. Additionally, the number of iterations (167) is decreased by a factor of four, demonstrating a higher level of efficiency. The remaining distance for the Fc-MPC (19.72 km) is shortened by 8.34 km, which significantly enhances accuracy. Fc-MPC achieves a mean path deviation of 0.20 km and a maximum path deviation of 0.30 km, demonstrating a more precise and consistent navigation route, as shown in \Cref{fig:9}.

\begin{table}[ht]
\centering
\fontsize{9}{9}\selectfont
\caption{Performance comparison of LTV-MPC and Fc-MPC under short-term magnetic storm interference.}
\resizebox{\columnwidth}{!}{%
\begin{tabularx}{\columnwidth}{lXX}
\toprule
\textbf{Metric} & \textbf{LTV-MPC} & \textbf{Fc-MPC} \\ \midrule
Signal-to-noise ratio (dB) & 9.93 & 10.06 \\
Iterations & 171 & 167 \\
Step Var (km\(^2\)) & 0.45 & 0.41 \\
Trajectory length (km) & 847.53 & 827.00 \\
Mean path deviation (km) & 0.27 & 0.20 \\
Max path deviation (km) & 0.44 & 0.30 \\
Remaining distance (km) & 28.06 & 19.72 \\
\bottomrule
\end{tabularx}%
}
\label{tab:7}
\end{table}

The LTV-MPC algorithm exhibits a wider range of iterations, as shown in \Cref{fig:8a}, spanning from 163 to 179, with a high median of 174.5 and a 75th percentile at 179. Although it demonstrates significant fluctuations, Fc-MPC shows a more concentrated range (162 to 170) with fewer iterations, indicating improved computational efficiency under short-term magnetic storm interference. As shown in \Cref{fig:8b}, the path variability of the LTV-MPC algorithm displays a wide range, from 0.31 km$^2$ to 0.74 km$^2$, with a high median of 0.47 km$^2$, reflecting poor stability. Conversely, the variability of the Fc-MPC algorithm is comparatively small, ranging from 0.28 km$^2$ to 0.59 km$^2$, with a low median of 0.38 km$^2$, indicating enhanced stability.

The trajectory length range for the LTV-MPC algorithm is extensive (807.35 km to 888.04 km), with a high median of 864.90 km and considerable fluctuation, as shown in the \Cref{fig:8c}. Conversely, the path length generated by the Fc-MPC algorithm is shorter and more concentrated (ranging from 781.29 km to 805.22 km), with a lower median of 793.09 km and greater stability. Consequently, the Fc-MPC algorithm demonstrates superior performance in navigation path planning. In the \Cref{fig:8d}, the LTV-MPC algorithm again shows considerable fluctuations in the distance from the actual arrival position to the destination, with values ranging from 22.30 km to 32.70 km and a higher median of 28.83 km, indicating relatively low positioning accuracy. Conversely, the distance distribution for the Fc-MPC algorithm is more concentrated, ranging from 15.35 km to 23.86 km, with a lower median of 18.77 km and fewer finite outliers, suggesting enhanced navigation stability.

In conclusion, the data related to the Fc-MPC algorithm is more tightly clustered, exhibiting a lower median and a reduced range between the maximum and minimum values. This suggests that the Fc-MPC algorithm demonstrates greater accuracy and stability, whether in the context of long-term or short-term magnetic storm interference. Furthermore, the Fc-MPC algorithm shows a lower incidence of finite outliers, which further supports its superior performance. Overall, the Fc-MPC algorithm is considered more effective than the LTV-MPC algorithm in terms of both accuracy and stability.
\subsection{Measured Experiment}
\label{Section:4.3}
\begin{figure}[htp]
    \centering
    \includegraphics[width=0.7\linewidth]{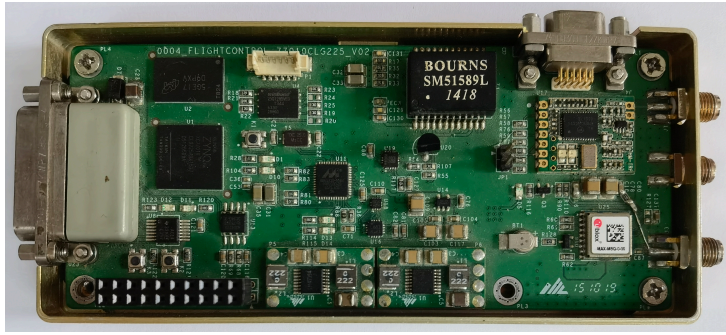}
    \caption{Combined navigation system equipped with GPS, inertial and magnetometer sensors.}
    \label{fig:10}
\end{figure}

In the experiments described in this section, the methods are evaluated by using actual measured data. The measurement device used in the study is shown in \Cref{fig:10}. The equipped ublox MAX M8Q provides GPS location data, MPU9250 provides magnetic field and inertial navigation data, and ZYNQ-7020 completes data processing and heading calculation. The long-distance data collection was conducted from Xi'an (34.43\(^\circ\)N, 108.75\(^\circ\)E) to Jiayuguan (39.86\(^\circ\)N, 98.34\(^\circ\)E). 

\begin{figure}[htp]
    \centering
    \includegraphics[width=0.7\linewidth]{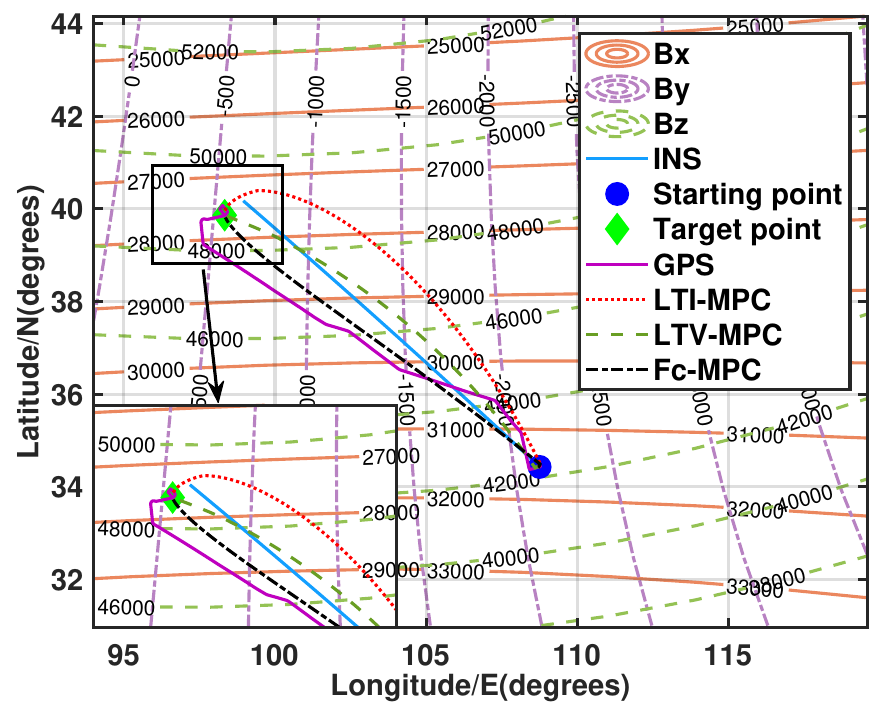}
    \caption{Navigation trajectories under measurement data.}
    \label{fig:11}
\end{figure}

In this section of the experiment, we performed trajectory planning and heading calculations using real measured magnetic and inertial navigation data. The results are recorded in \Cref{tab:8}, and the corresponding trajectory outcomes obtained through Fc-MPC, LTI-MPC, LTV-MPC, INS, and the recorded GPS trajectory are plotted in \Cref{fig:11}.

\begin{table*}[htp]
\centering
\fontsize{9}{9}\selectfont
\caption{Navigation performance evaluation under real measurement data.}
\label{tab:8}
\resizebox{\columnwidth}{!}{%
\begin{tabular}{@{}lccccccccc@{}}
\toprule
\textbf{} & \textbf{Itrations} & \textbf{\begin{tabular}[c]{@{}c@{}}Step Var\\ (km\(^2\))\end{tabular}} & \textbf{\begin{tabular}[c]{@{}c@{}}Trajectory\\length (km)\end{tabular}} & \textbf{\begin{tabular}[c]{@{}c@{}}Mean path \\ deviation (km)\end{tabular}} & \textbf{\begin{tabular}[c]{@{}c@{}}Maximum path \\ deviation (km)\end{tabular}} & \textbf{\begin{tabular}[c]{@{}c@{}}PMR\\(\%)\end{tabular}} & \textbf{\begin{tabular}[c]{@{}c@{}}CEP\\(km)\end{tabular}} & \textbf{\begin{tabular}[c]{@{}c@{}}Latitude \\ (N\(^\circ\))\end{tabular}} & \textbf{\begin{tabular}[c]{@{}c@{}}Longitude \\ (E\(^\circ\))\end{tabular}} \\ \midrule
GPS & 867 & 31.43 & 1367.56 & 0.45 & 0.83 & 34.60 & 0.00 & 98.34 & 39.86 \\
INS & 268 & 3.25 & 1072.10 & 0.19 & 0.69 & 78.36 & 62.71 & 98.95 & 40.17 \\
LTI-MPC & 244 & 32.42 & 1202.99 & 1.08 & 1.56 & 10.25 & 9.17 & 98.34 & 39.94 \\
LTV-MPC & 224 & 752.41 & 1114.07 & 0.34 & 0.52 & 45.09 & 6.53 & 98.41 & 39.84 \\
Fc-MPC & 226 & 35.33 & 1114.00 & 0.18 & 0.33 & 86.37 & 5.67 & 98.36 & 39.81 \\ \bottomrule
\end{tabular}%
}
\end{table*}

To address the problem of subtle gradient variation, the Fc-MPC algorithm is proposed based on the LTV-MPC algorithm. As shown in \Cref{tab:8}, Fc-MPC exhibits not only a reduced number of iterations (226) but also a shorter navigation path (1114.0 km). Compared to LTI-MPC and LTV-MPC algorithms, Fc-MPC significantly enhances navigation efficiency by reducing the navigation trajectory by 0.07 km to 253.56 km. Remaining distance is a crucial indicator that reflects the precision of the navigation path; Fc-MPC (5.67 km) is notably smaller compared to other algorithms. These performance analyses indicate that the Fc-MPC algorithm introduced has the potential to enhance navigation efficiency, stability, and precision.

Under the simulation and experimental analysis presented in this paper, several areas for optimization and exploration in future research have been identified. This study bases its reliability on the assumption that a single sensor returns measurement data during the navigation process. Additionally, this paper ignores the measurement deviations caused by the magnetization of the geomagnetic field on the carrier during the simulation process, which may result in decreased accuracy in practical applications. Future research could focus on developing a sensor matrix and employing multi-sensor data fusion technology to enhance the accuracy and reliability of measurement outcomes.

\section{Conclusion}
\label{Section:5}
This paper proposes the Fc-MPC algorithm to enhance the precision, efficiency, and stability of geomagnetic and inertial combined navigation systems. In GPS-denied environments without prior geomagnetic maps, the Kalman filter is employed to fuse the output of the inertial navigation error equation with the position error derived from geomagnetic navigation. The Fc-MPC algorithm adds an error correction method to each step of the MPC prediction results. This solves the cumulative error problem that comes with inertial navigation systems and the effect that magnetic field interference has on geomagnetic navigation systems. Simulations and actual data experiments demonstrated that the Fc-MPC method can successfully complete navigation missions in environments without geomagnetic maps, also exhibits strong performance in magnetic storm conditions.

%%       \citet{<label>} ==> Lamport (1994)
%%       \citep{<label>} ==> (Lamport, 1994)
%%




% %% For authoryear reference style

% \bibitem[Lamport(1994)]{lamport94}
%   \textit{\LaTeX: a document preparation system},
%   Addison Wesley, Massachusetts,

% \end{thebibliography}

%\bibliographystyle{cas-model2-names}

\bibliographystyle{elsarticle-harv}
\bibliography{cas-refs}

\end{document}